\documentclass[structabstract]{aa}  
\usepackage{graphicx}
\usepackage{txfonts}
\usepackage{sidecap}
\usepackage{rotating}
\usepackage{natbib}
\usepackage{longtable}
\bibpunct{(}{)}{;}{a}{}{,}
\def\Ha{H$_{\alpha}$}
\def\Hb{H$_{\beta}$}
\def\Hg{H$_{\gamma}$}
\def\Hd{H$_{\delta}$}
\def\O3{O\,{\sc iii}}
\def\S2{S\,{\sc ii}}
\def\N2{N\,{\sc ii}}

\def\HeI{He\,{\sc i}}
\def\FeI{Fe\,{\sc i}}
\def\FeII{Fe\,{\sc ii}}

\def\MgII{Mg\,{\sc ii}}

\def\CaII{Ca\,{\sc ii}}
\def\sun{$_{\odot}$}
\def\x{$\times$}

\begin{document}

   \title{Var~C: Long-term photometric and spectral variability\\
          of an LBV in M33
       \thanks{Based on observations collected at the Th\"uringer
         Landessternwarte (TLS) Tautenburg.}
       \thanks{Based on observations collected at the Centro Astron\'{o}mico
         Hispano Alem\'{a}n (CAHA) at Calar Alto, operated jointly by the
         Max-Planck Institut f\"ur Astronomie and the Instituto de
         Astrof\'{i}sica de Andaluc\'{i}a (CSIC).}\fnmsep
   }

   \subtitle{}

   \author{B.~Burggraf
          \inst{1}
          \and
          K.~Weis
          \inst{1}
          \and
          D.~J.~Bomans
          \inst{1}
          \and
          M.~Henze
          \inst{2}\fnmsep\inst{3}
          \and
          H.~Meusinger
          \inst{3}
          \and
          O.~Sholukhova
          \inst{4}
          \and \\
          A.~Zharova
          \inst{5}
          \and
          A.~Pellerin
          \inst{6}
          \and
          A.~Becker
          \inst{1}
   }

   \institute{Astronomisches Institut der Ruhr--Universit\"at Bochum, 44780 Bochum, Germany\\
              \email{burggraf@astro.rub.de}
          \and
              Max-Planck-Institut f\"ur extraterrestrische Physik, 85748 Garching, Germany
          \and
              Th\"uringer Landessternwarte Tautenburg, 07778 Tautenburg, Germany
          \and
              Special Astrophysical Observatory, Russian Academy of Sciences, Nizhnij Arkhyz 369167, Russia
          \and
              Lomonosov Moscow State University, Sternberg Astronomical Institute, 13 Universitetskij prospekt, Moscow 119234, Russia
          \and
              Department of Physics and Astronomy, SUNY-Geneseo, 1 College Circle, Geneseo, NY 14454, USA
   }

   \date{}

% \abstract{}{}{}{}{} 
% 5 {} token are mandatory
 
  \abstract
   {}
   {So far the highly unstable phase of luminous blue variables (LBVs)
     has not been understood well. It is still uncertain why and which massive
     stars enter 
     this phase. Investigating the variabilities by looking for a possible
     regular or even (semi-)periodic behaviour could give a hint at the
     underlying mechanism for these variations and might answer the question
      of where these variabilities originate.
     Finding out more about the LBV phase also means understanding
     massive stars better in general, which have (e.g. by enriching the ISM with
     heavy elements, providing ionising radiation and kinetic energy) a
     strong and significant influence on the ISM, hence also on their host 
     galaxy.}
   {Photometric and spectroscopic data were taken for the LBV Var~C in M33 to
     investigate its recent status. In addition, scanned historic plates, 
     archival data, and data from
     the literature were gathered to trace Var~C's behaviour in the
     past. Its long-term variability and periodicity was
     investigated.}
   {Our investigation of the variability indicates possible (semi-)periodic
     behaviour with a period of 42.3~years for Var~C. That Var~C's light curve
     covers a time span of more than 100~years means that more than two full periods
       of the cycle are visible. The critical historic maximum around 1905 is
       less strong but discernible even with the currently rare historic
       data. The semi-periodic and secular structure of the light curve is
       similar to the one of LMC R71. Both light curves hint at a new aspect
       in the evolution of LBVs.}
   {}


   \maketitle
%
%________________________________________________________________

\section{Introduction}

Luminous blue variables (LBVs) are stars in a short phase that lasts only
several 10$^{4}$~years \citep[e.g. ][]{1994PASP..106.1025H} towards the end of
the evolution of some of the most massive and most luminous stars, between their
main sequence and Wolf-Rayet (WR) phase. The initial masses of LBVs
  covered the range of 50 to 120~M\sun. Newer models, however, that include
  rotation can also reproduce LBV progenitor stars with masses as low as
21~M\sun\ \citep{2005AA...429..581M}, matching the observations. LBVs have
luminosities of about 10$^{6}$~L\sun.

\onlfig{1}{
\begin{figure*}
   \centering
   \includegraphics[width=\textwidth]{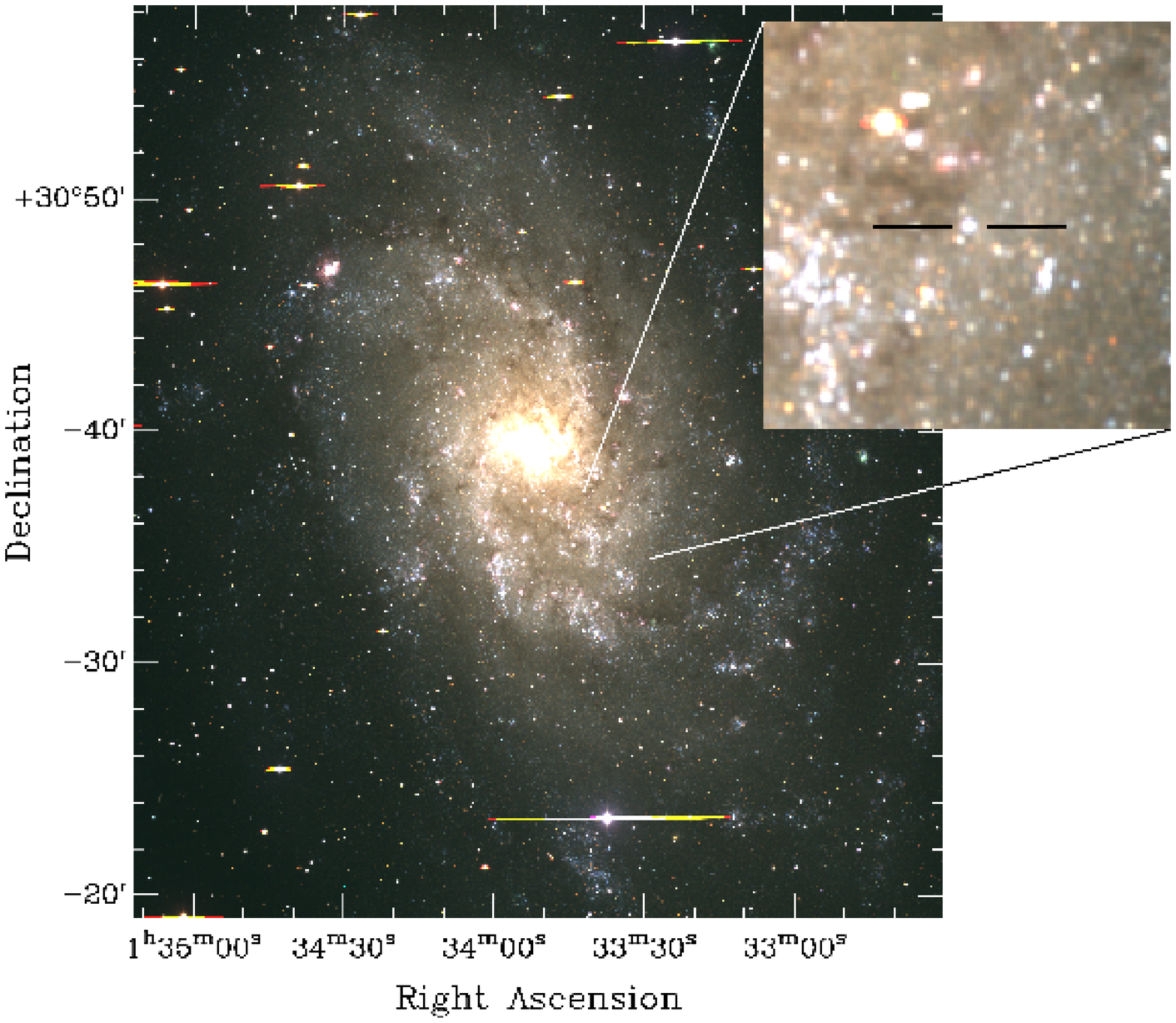}
   \caption{RGB-image of M33 and Var~C. The colour image was generated from
     the R, V, and B band images from the Th\"uringer Landessternwarte (TLS)
     Tautenburg taken 6 January 2008. North is up and east to the left.}
   \label{image_rgb}
\end{figure*}
}

An important property of LBVs -- one that first defines them as LBVs -- is their
variability. LBVs show variability on different timescales (months, years, or
decades) and with different amplitudes (a tenth of magnitudes up to $>$
2~mag) \citep{1994PASP..106.1025H}. A variability intrinsic to LBVs is the
so-called S~Dor variability. It occurs on a timescale of about 10-40~years
during which the visual brightness rises 1-2 magnitudes while the bolometric
brightness remains nearly constant. A more subtle distinction by
\citet{2001AA...366..508V} is the classification of long ($>$ 20~years, L-SD)
and short ($<$ 10~years, S-SD) S~Dor variability, and the ex-/dormant class,
that contains all LBVs that have not been active on a longer---not further
defined---timescale.

These variations can be superimposed.
The S~Dor variability -- also known as S~Dor cycle or S~Dor eruption -- must
be distinguished from so-called giant eruptions. While undergoing a giant
eruption, LBVs can show even larger photometric variations of more than
two magnitudes \citep{1994PASP..106.1025H}. A prominent example
for them is the giant eruption of the LBV $\eta$~Car in the 19th
century. Giant eruptions of LBVs are also important because they might be
mistaken for supernova explosions \citep{2005AA...429L..13W}. So far it is
unknown whether such giant eruptions can occur more than once in the same
LBV.
Therefore, a photometric monitoring and analysis is one of the -- maybe 
the best -- method(s) of actually pinpointing an LBV. 
In particular, to find 
and disentangle candidates in the L-SD and ex-/dormant classes 
(see above), establishing and analysing long-term light curves are
essential. At the same time, this increases the chances of catching an LBV,
for the first time, with  a giant eruption and a known photometric post
eruption history. Last but not least, long-term light curves can be checked
for regular and periodic changes. For short-term S~Dor LBVs, 
periodicities in the light curve have already been reported by
\citet{2001AA...366..508V}. With only very few light curves spanning a large
time span (e.g. 50~years and more), an analysis for the long-term S~Dor LBVs is
still missing. With the addition of the light curve for Var~C presented here
and a recently reported long-term study of R71 (see Section 6), 
this kind of study starts to unfold.

LBVs not only show photometric variability, but also reveal different spectra
depending on the state they are in. In their quiet -- minimum light -- phase,
LBVs show the spectrum of a hot supergiant. Prominent are H, He, \FeII,\ and
[\FeII] lines in emission, often also with P-Cygni profiles. At a phase of
maximum light, the spectrum resembles that of a cool A -- F type star
\citep{1994PASP..106.1025H}. These changes in the spectral type can be
associated with changes in the star's apparent temperature. The corresponding
shift of the Planck curve, hence of the colour of the star, is the
mechanism behind the S~Dor variability of an LBV.

LBVs have a high mass loss rate. During their quiet phase, this mass loss rate
is of the order of 10$^{-7}$ -- 10$^{-5}$~M\sun/yr, while 
during their S~Dor eruption phase, this can reach up to 
10$^{-5}$ -- 10$^{-4}$~M\sun/yr \citep{1994PASP..106.1025H}.

\begin{figure}
   \centering
   \includegraphics[width=\columnwidth]{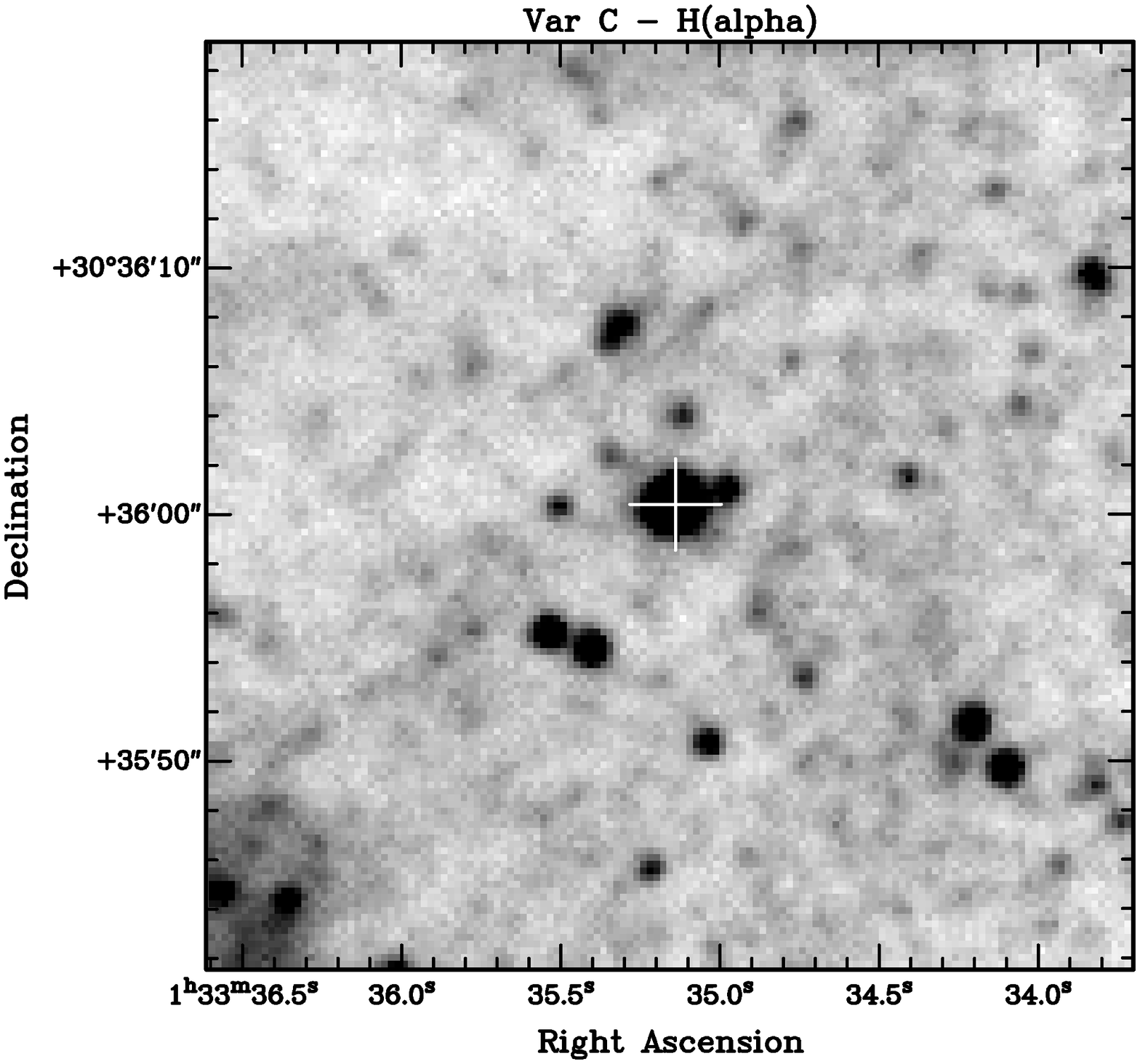}
   \includegraphics[width=\columnwidth]{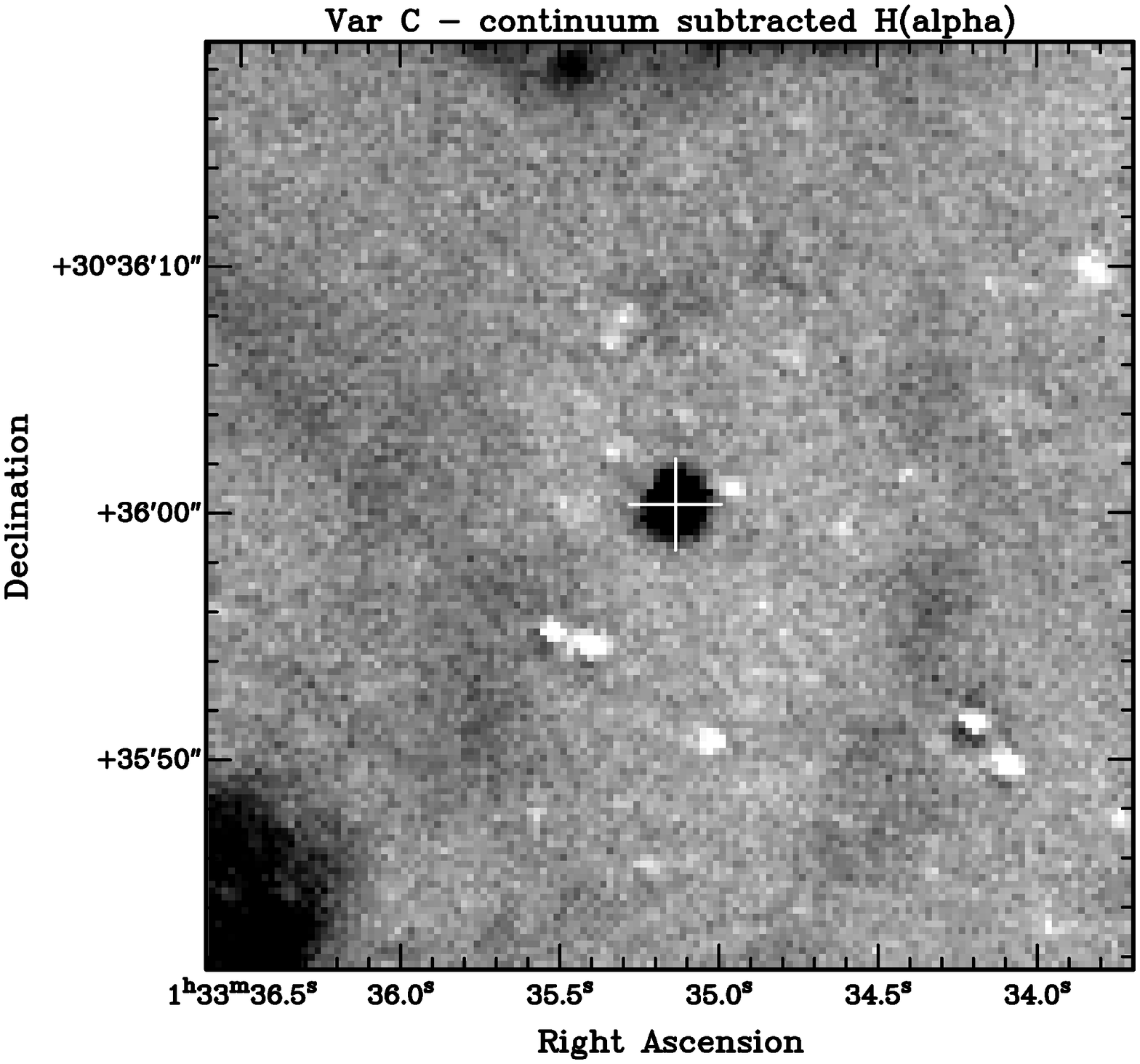}
   \caption{\Ha -image (upper panel) and continuum-subtracted \Ha -image (\Ha
     -R; lower panel) of Var~C and its surroundings. Images were produced from
     NOAO LGGS data and have a size of about 150\,pc\,$\times$\,150\,pc. North
     is up and east is to the left.}
   \label{image}
\end{figure}

By sweeping up the slower older stellar wind of earlier phases and S~Dor
cycles, as well as through mass ejections during a giant eruption, small
\citep[$<$ 5 pc, ][]{2011BSRSL..80..440W} circumstellar LBV nebulae are
formed. These LBV nebulae contain CNO processed material from the star and are
therefore rich in nitrogen and helium but low in oxygen and carbon
\citep{1983AA...120..113M}.

To further determine the characteristics of LBVs it is necessary to find more
of these stars. In Local Group galaxies, stars can be resolved spatially, 
and individual stellar parameters can be derived. The Local Group also 
provides different types of galaxies, hence
different environments, and is therefore an ideal place to look for LBVs.

Periodicity on smaller timescales has already been known to occur in 
LBVs. The photometric variations in AG~Car, an LBV in the Milky Way,
where found by \citet{2001AA...366..508V} to oscillate with periods of the
order of one year (P$_0$=371.4 d; P$_1$=305 d or 475 d). Both periods appear
superimposed. \citet{2000MNRAS.311..698S} reported the M33 candidate LBV
UIT301 (also named B416) to show a periodicity of 8.26 days.

The periodicities found so far belong to the short S~Dor variability
(S-SD). Only a few indications for periodicities of the long S~Dor 
type (L-SD) have been reported \citep{1997A&A...318...81V}.
A likely reason for it is that for most LBVs, the time baseline of the light
curve is too short. Even if data from past centuries are available, these are
often only a few data points so that the resulting light curve is quite patchy.
A recently reported exception is the LMC LBV R71 \citep{2014ATel.6295....1W}.

The Local Group spiral galaxy M33 (classified as Scd) is located at a distance
of approximately 840\,kpc \citep[distance modulus of 24.53$\pm$0.11 mag,
][]{2009MNRAS.396.1287S}. 
The proximity and low inclination of M33 make it an excellent target for
studying single stars. It also provides the possibility to study LBVs in an
environment with a fairly low metallicity (Z=0.008).

\citet{1953ApJ...118..353H} first noticed the variability of 
several massive stars, which include Var~C in M33. Based on this publication, 
those stars were referred to as Hubble--Sandage variables. Finally 
together with the P~Cyg- and S~Dor-like variables, they were united 
under the term LBVs \citep{1984IAUS..105..233C}.

\begin{table}
\caption{Log of observations.}
\label{observationlog}
\centering
\begin{tabular}{l r l l}
\hline\hline
Date                     &            & Filter   & Site \\
\hline
13, 31                   & Aug.  2002 & B, V     & WIYN \\
14                       & Sept. 2002 & B, V     & WIYN \\
10                       & Oct.  2002 & B, V     & WIYN \\
11                       & Nov.  2002 & B, V     & WIYN \\
 7, 23                   & Jan.  2003 & B, V     & WIYN \\
19, 26                   & Sept. 2003 & B, V     & WIYN \\
20                       & Oct.  2003 & B, V     & WIYN \\
15                       & Nov.  2003 & B, V     & WIYN \\
27                       & Dec.  2003 & B, V     & WIYN \\
16                       & Jul.  2005 & B, V     & WIYN \\
 8, 9, 10, 12, 13        & Oct.  2005 & V        & COoSAI \\
28                       & Jan.  2006 & B        & TLS \\
27                       & Jul.  2006 & B        & CAHA \\
 4                       & Sept. 2006 & B        & CAHA \\
 2, 5, 8, 10, 11, 13, 14 & Oct.  2006 & V        & COoSAI \\
22                       & Oct.  2006 & B        & CAHA \\
 2                       & Oct.  2007 & B, V     & COoSAI \\
 3                       & Oct.  2007 & V        & COoSAI \\
11                       & Oct.  2007 & B, V     & COoSAI \\
14                       & Oct.  2007 & B, V     & TLS \\
15, 17                   & Oct.  2007 & B, V     & COoSAI \\
 6                       & Jan.  2008 & B, V     & TLS \\
 2                       & Oct.  2008 & B, V     & TLS \\
22                       & Dec.  2008 & B, V     & TLS \\
20                       & Jan.  2009 & B, V     & TLS \\
28                       & Feb.  2009 & B, V     & TLS \\
19                       & Sept  2009 & B        & TLS \\
19                       & Oct.  2009 & B        & TLS \\
21, 22                   & Oct.  2009 & B, V     & TLS \\
 5, 8                    & Nov.  2009 & B, V     & COoSAI \\
20                       & Feb.  2010 & B, V     & TLS \\
25                       & Feb.  2011 & B        & TLS \\
 2                       & Mar.  2011 & V        & TLS \\
31                       & Aug.  2011 & B        & TLS \\
 2                       & Sept. 2011 & B        & TLS  \\
 4                       & Sept. 2011 & V        & TLS \\
29                       & Sept. 2011 & B        & TLS \\
30                       & Sept. 2011 & V        & TLS \\
 1, 3                    & Oct.  2011 & B, V     & TLS \\
\hline
\end{tabular}
\end{table}

With $\alpha_{2000}$=1:33:35.14 and $\delta_{2000}$=30:36:00.55, Var~C is
located only about 5\arcmin\ south-west of the centre of M33 (Figure
\ref{image_rgb}). Figure \ref{image} shows the \Ha -image (upper panel) and
the continuum-subtracted \Ha -image (lower panel) of Var~C and its
surroundings. After comparing both images, Var~C shows a bright \Ha\ emission,
while neighbouring stars do not.

In the continuum-subtracted \Ha -image, Var~C seems to be surrounded by a faint
ring-nebular-like structure with a radius of approximately 50 pc, which would
be consistent with the star's main sequence bubble. \cite{1977ApJ...218..377W}
give a radius of approximately 30 pc for the size of such a bubble, depending
on the density of the surrounding interstellar medium (ISM).

Since both photometric and spectral observations of Var~C have been made
numerous times in the past, this provides us with the rare chance to trace the
behaviour of an LBV over a wide span of time. For Var~C we compiled a light
curve that has a significant length (more than 100~years) in combination with
a reasonable coverage of data points. This gives the chance not only to look
for long time variations, but also to check whether these variations might be
periodic. Since the origin of the variabilities in LBVs is not known yet,
finding a periodic behaviour could give a hint of the underlying mechanism for
these variations.

%__________________________________________________________________

\section{Observations and data analysis}

\subsection{CCD Imaging}

\subsubsection{Observations from TLS}

Observations with the 2m-telescope of the {\it Th\"uringer Landessternwarte}
(TLS) Tautenburg in its Schmidt mode were taken during several runs between
January 2006 and October 2011. The SITe detector was used with 2048\x 2048
pixels and an image scale of 1\farcs23/pixel (FOV: 42\arcmin \x 42\arcmin) in
combination with broadband Johnson B, V, R filters or a narrowband (100\AA)
\Ha\ filter, respectively. A standard data reduction (bias correction,
flatfielding) was carried out using IRAF and a photometric analysis was
performed using CCDCAP \citep{ccdcap}.

\subsubsection{Observations from CAHA}

We derived photometric data during our M33 survey using the {\it Calar Alto
  Faint Object Spectrograph} (CAFOS) in its imaging mode at the 2.2m-telescope
at the Calar Alto Observatory ({\it Centro Astron\'{o}mico Hispano Alem\'{a}n}
(CAHA)). The data were recorded from July to October 2006 during three runs in
service mode using CAFOS at the 2.2m-telescope. SITe-1d detector with 2048\x
2048 pixels, an image scale of 0\farcs53/pixel and a field of view of about
16\arcmin \x 16\arcmin\ was used. Images were taken in B and R filters. Basic
data reduction was performed using IRAF. Subsequently, photometry was carried
out using DOLPHOT \citep{2000PASP..112.1383D}.

\subsubsection{Observations from COoSAI}

CCD observations at the {\it Crimean Observatory of Sternberg Astronomical
Institute} (COoSAI) were made during 19 nights between October 2005 and
October 2009. The 60cm reflector (C60) equipped with an Apogee AP47
camera and a wide-field telescope (the 50cm Maksutov camera-AZT-5) equipped
with a Pictor camera were used. The typical uncertainty of our estimates is
0.1~mag (0.2-0.5~mag for stars fainter than 17 mag).

\subsubsection{Observations from WIYN}

Between August 2002 and July 2005, observations were made with the
3.5m-telescope at the {\it Wisconsin-Indiana-Yale-NOAO} (WIYN)
Observatory. The Mini-Mosaic imager consisting of two CCDs with 4096\x 2048
pixels, a field of view of 9\arcmin 6\x 9\arcmin 6, and an effective sampling
of 0\farcs28/pixel was used. A basic data reduction was performed using
IRAF. PSF photometry was carried out using IRAF/DAOPHOT. A detailed
description of the analysis method is given in \cite{2011ApJS..193...26P}.

\subsubsection{Observations from the DIRECT project}

Unpublished data from the DIRECT project taken between September 1996 and
November 1998 were used to supplement the light curve. A description of the
data and the analysis method can be found in \cite{2011ApJS..193...26P}.
The DIRECT project \citep[e.g. ][]{1998AJ....115.1016K} was carried out
to determine the direct distances to M31 and M33 using Cepheids and detached
eclipsing binaries. CCD observations with 1-m class telescopes in BVI
filters started in 1996.

\subsubsection{Archival data from NOAO}

Archival data from the {\it National Optical Astronomy Observatory
 Local Group Galaxies Survey} (NOAO LGGS) \citep{2001AAS...19913005M} taken
with the 4m-telescope at {\it Kitt Peak National Observatory} (KPNO) in
October 2000 and September 2001 were also used to perform photometric analysis. These images consisted of 8192\x 8192 pixels and had an image scale of
0\farcs271/pixel (FOV: 36\arcmin \x 36\arcmin).

A basic data reduction (bias correction, flat fielding, etc.) was already done
by \citet{2002AAS...20110407M}, but photometry was not yet available when we
started our project. We therefore performed PSF photometry using
IRAF/DAOPHOT. The comparison with \citet{2006AJ....131.2478M} revealed only a
small photometric offset of a tenth of a magnitude, which was corrected.

\subsection{Photographic plate scans from TLS}

Our photometric data set was supplemented by archival photographic plates of
M33 taken between August 1963 and December 1996 with the Tautenburg Schmidt
telescope (free aperture 1.34m, focal length 4m). A single Tautenburg plate
covers an unvignetted field of $3\,\fdg3 \times 3\,\fdg3$ with a plate scale
of $51\,\farcs4$ per mm. We made use of 77 B plates, all of which had been
digitised with the Tautenburg Plate Scanner \citep{1999AAS..139..141B}. The
digital images have a pixel size of $0\,\farcs5 \times 0\,\farcs5$.

Data was reduced in an analogous way to the method applied by
\citet{2008AA...477...67H} for analysing digitised photographic plates of M31
taken with the same telescope. We used the Source Extractor package
\citep{1996A&AS..117..393B} for object detection, background correction, and
relative photometry. The photometric calibration was done using the M33 part
of the Local Group Galaxies Survey \citep[LGGS,][]{2006AJ....131.2478M}. A
detailed description of the analysis method is given in
\citet{2008AA...477...67H}.

\subsection{Heidelberg Digitized Astronomical Plates}

We used the {\it Heidelberg Digitized Astronomical Plates} (HDAP) to
obtain additional datapoints for the sparsely sampled era at the beginning
of the 20th century. Twelve photographic plates of M33 were taken with the
72cm Walz-Reflector between 1908 and 1922. No detailed or specific information
on the spectral sensitivity of the emulsions that were used could be derived
from the observers' notebook, except for one case (``blue sensitive''). We
therefore treated all plates as blue-sensitive ``normal'' plates and calibrated
the magnitudes into B band. We measured the brightness of Var~C to fill the
sparsely populated light curve for those epochs. The large spread in the
quality of the  plates prevented a consistent linearisation for this
dataset. Therefore, magnitudes were derived using the classical Argelander
method \citep{1901AN....154..413N}.

\subsection{Historical data taken from literature}

To trace the behaviour of Var~C in the past, a light curve was created using
data from both the literature and the archive. Since the historical data were
either photographic magnitudes ($m_{pg}$) or B magnitudes, we compiled a light
curve in B. We calculated $m_{pg}$ into Johnson B magnitudes using the following
conversion given in \citet{2010AA...512A...1M}:

\begin{equation}
    B = m_{pg} + 0.1
.\end{equation}

\noindent
B magnitudes were calculated for data points taken from
\citet{1953ApJ...118..353H}, \citet{1973PZ.....19....3S},
\citet{1973AA....22..453R}, \citet{1988IBVS.3193....1L}, and
\citet{1990SvA....34..364S}.

In case no tabulated data existed, $m_{pg}$ data points were extracted from
given light curves using DEXTER \citep{2001ASPC..238..321D}. Errors in time
and magnitude are mainly due to uncertainties from extracting the data. A
maximum error was assumed to be half the size of the data point symbol in the
light curve plot. This gave an error of JD$\pm$36 days and $m_{pg}\pm$0.051 mag
for data values taken from \citet{1953ApJ...118..353H}, JD$\pm$18 days and
$m_{pg}\pm$0.023~mag for values from \citet{1973AA....22..453R}, and
JD$\pm$220 days and B$\pm$0.094~mag for values from
\citet{2004CoAst.145...28Z}.

Magnitudes derived from photographic plates or photometers already given in B
magnitudes were taken from \citet{1975ApJS...29..303V},
\citet{1978ApJ...219..445H}, \citet{1978ApJ...221L..73H},
\citet{1984ApJ...278..124H}, \citet{1988AA...203..306H}, and
\citet{1999AA...349..796K}.
Further B CCD-photometry values came from \citet{1991AJ....101.1663W},
\citet{1995AJ....110.2715M}, \citet{1996AA...314..131S},
\citet{2001AJ....122.2477M}, \citet{2006AA...458..225V},
\citet{2013ATel..5362....1H}, and \citet{2013ATel.5538....1V}.

All B data values are collected in Tables \ref{Lightcurve_data_literature}
and \ref{Lightcurve_data}, given in the appendix.

\begin{figure*}
   \centering
   \includegraphics[width=\textwidth]{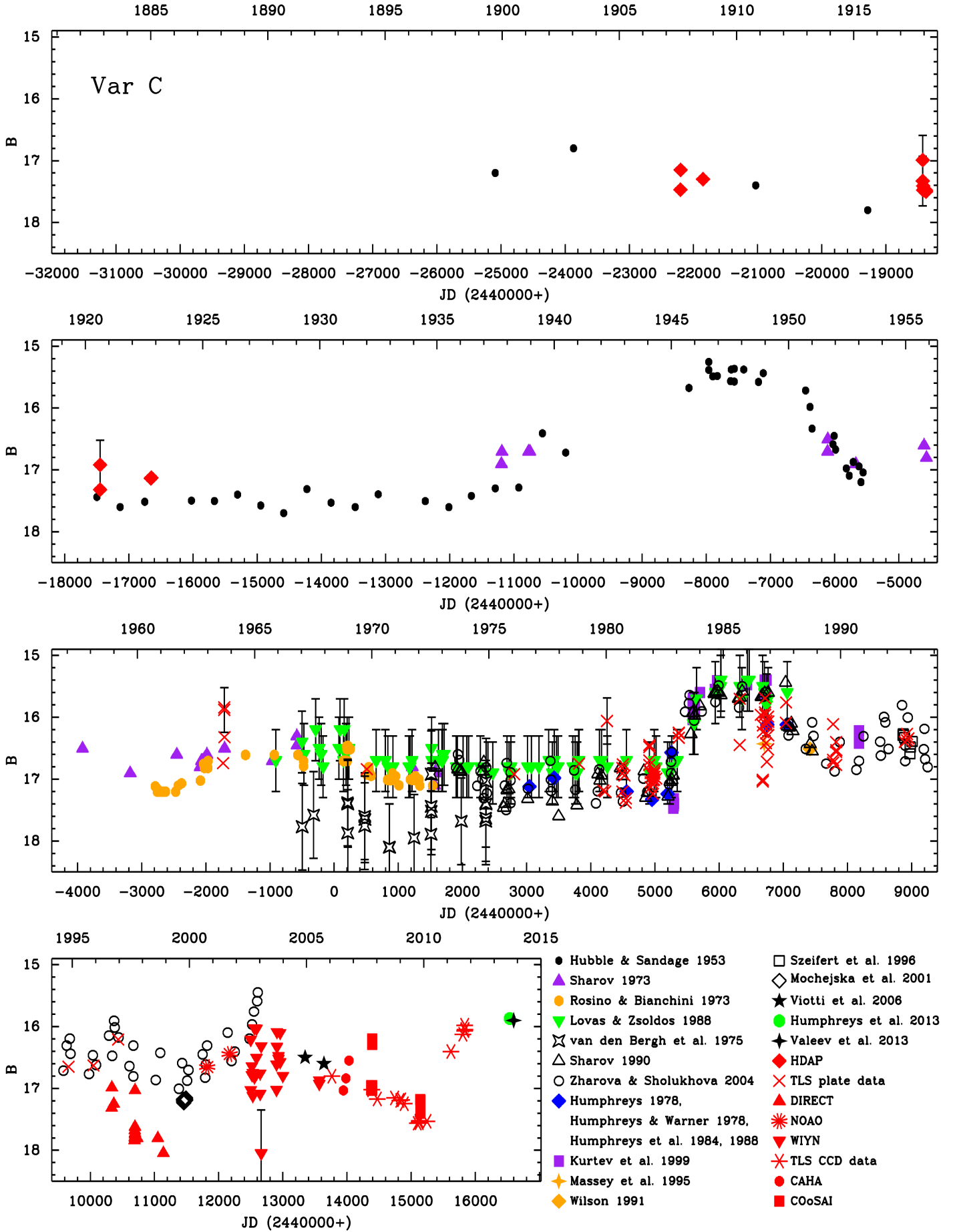}
      \caption{B light curve of Var~C between 1899 and 2013. For reasons
          of clarity and readability error bars are only given for data with
          uncertainties of more than 0.4~mag. The data values are listed in
        Table  \ref{Lightcurve_data}. Where available, also magnitude errors
        are given there.}
      \label{licu_VarC}
\end{figure*}

Additional V band data used in Figure \ref{licu_BV} are from
\citet{Spiller92}, \citet{2001AJ....121..870M}, and
\citet{2006MNRAS.370.1429S}.

For completeness, in the following we list published values for Var~C, which
had no observation times given. Therefore, these values could not be included
in the light curves. The catalogue from \citet{1993ApJS...89...85I} lists
Var~C as star IFM\_B~600. Values of V=15.20 mag, B$-$V=0.30~mag (B=15.50 mag),
and U$-$B=$-$0.40~mag (U=15.10 mag) are given. Var~C is listed as v267 by
\cite{1995ApSS.226..229F} with values of V=16.70~mag and B$-$V=0.30 mag
(B=17.00 mag).

\subsection{Spectral observations from SAO RAS}

Two spectra were taken on 13 November 2004 and 22 October 2008 with the
6m-telescope at the {\it Special Astrophysical Observatory of the Russian
Academy of Sciences} (SAO RAS). The {\it Multi-Pupil Fiber Spectrograph}
(MPFS) \citep{2007ASPC..361..491S} and the {\it Spectral Camera with Optical
Reducer for Photometrical and Interferometrical Observations} (SCORPIO)
\citep{2005AstL...31..194A} were used respectively.

%__________________________________________________________________

\section{Light curve of Var~C}

\onltab{2}{
\begin{table*}
\caption{Summary of optical spectra of Var~C.}
\label{Spec_param}
\centering
\begin{tabular}{l l l l c}
\hline\hline
Date & JD & Features & Spectral Type & Ref. \\
\hline
30 Nov. 1946                    & 2432155            & \CaII\ K and \CaII\ H in absorption with               & later than F0        & 1 \\
                                &                    & nearly equal strength                                  &                      &   \\
                                &                    & \Hg\ and \Hd\ weakly in absorption                     &                      &   \\
23 Aug. 1947                    & 2432421            & \CaII\ K and \CaII\ H in absorption with               & later than F0        & 1 \\
                                &                    & nearly equal strength                                  &                      &   \\
24/25 July -- 22 Sept. 1949     & 2433122 -- 2433182 & faint \Hb\ in emission                                 &                      & 1 \\
3 Aug. -- 28/31 Oct. 1951       & 2433862 -- 2433951 & \Hb\ and \Hg\ in emission                              & late A-type\tablefootmark{a} & 1 \\
                                &                    & \CaII\ K and \CaII\ H in absorption                    &                      &   \\
29 Nov. 1951                    & 2433980            & \Hb\ and \Hg\ in emission                              &                      & 1 \\
27/28 Sept. 1973                & 2441953            & bright \Ha\ in emission                                &                      & 2 \\
14 Oct. -- 10 Nov. 1974         & 2442335 -- 2442362 & several hydrogen, \HeI\, \FeII,                        & late B-type$^a$      & 3 \\
                                &                    & and [\FeII] lines in emission                          &                      &   \\
                                &                    & \CaII\ K in absorption                                 &                      &   \\
Aug. 1976                       & $\approx$2443006   & compared to previous spectrum:                         & early B-type$^a$     & 4 \\
                                &                    & hydrogen and \HeI\ emission lines weaker               &                      &   \\
                                &                    & some of the \FeII\ and [\FeII] lines no longer seen    &                      &   \\
10/11 Nov. 1983                 & 2445649.5          & \Ha, \Hg\ and numerous \FeII\ lines in emission        & A-type$^a$           & 5 \\
                                &                    & all with P-Cygni profiles                              &                      &   \\
                                &                    & strong \MgII\ $\lambda$4481 in absorption              &                      &   \\
Nov. -- Dec. 1985               & $\approx$2446385 -- $\approx$2446415 & \Ha\ and \Hb\ in emission            & F0Ia--F5Ia           & 6 \\
                                &                    & several strong \FeII\ and                              &                      &   \\
                                &                    & ionised metal lines in absorption                      &                      &   \\
                                &                    & I(\CaII\ K) $>$ I(\CaII\ H + H$_{\epsilon}$)            &                      &   \\
Aug. -- Sept. 1986              & $\approx$2446658 -- $\approx$2446689 & weaker metallic lines                & A2--A3               & 6 \\
                                &                    & I(\CaII\ H + H$_{\epsilon}$) $>$ I(\CaII\ K)            &                      &   \\
15/16/17 Sept. 1987             & 2447054 -- 2447056 & \FeII\ and other metallic lines in absorption          & A-type               & 6 \\
                                &                    & several \FeII-lines in emission                        &                      &   \\
                                &                    & \Ha, \Hb, and \Hg\ showed P-Cygni profiles             &                      &   \\
3/6 Oct. -- 13 Dec. 1992        & 2448899 -- 2448970 & weak \HeI\ lines in absorption                         & late B-type$^a$      & 7 \\
                                &                    & compared to previous spectra:                          &                      &   \\
                                &                    & hydrogen and \FeII\ emission lines getting stronger    &                      &   \\
14/15 Dec. 1993                 & 2449336.5          & pronounced hydrogen lines in emission                  & late B-type$^a$      & 8 \\
                                &                    & several \FeII\ and [\FeII] lines in emission           &                      &   \\
                                &                    & weak \HeI\ line                                        &                      &   \\
30 Nov. 2003                    & 2452974            & \Hb\ in emission and numerous metallic lines           & F0Ia--F5Ia           & 9 \\
                                &                    & in absorption                                          &                      &   \\
13 Nov. 2004                    & 2453323            & \Ha, \Hb, and \Hg\ in emission, some weak \FeI\        & A-type               & 10  \\
                                &                    & lines in absorption or emission                        &                      &   \\
7/8 Dec. 2004 -- 16 Jan. 2005   & 2453347 -- 2453387 & strong hydrogen lines in emission                      & B[e]                 & 11 \\
                                &                    & several \FeII\ and [\FeII] lines in emission           &                      &   \\
22 Oct. 2008                    & 2454762            & \Ha\ and \Hb, in emission, some \HeI\ lines and        & B-type               & 10  \\
                                &                    & several \FeI\ lines in emission                        &                      &   \\
29 Sept. -- 2 Oct. 2010         & 2455469 -- 2455472 & \Ha, \Hb, and \Hg\ in emission, several \HeI\ and      & B-type$^a$           & 9 \\
                                &                    & \FeII\ lines in emission, P-Cygni profiles             &                      &   \\
3 Oct. 2010                     & 2455473            & hydrogen, \HeI, \FeII, and [\FeII] lines in emission,  & B1--B2               & 12 \\
                                &                    & P-Cygni profiles                                       &                      &   \\
18 Sept. 2013                   & 2456554            & hydrogen lines in emission                             & A-type$^a$           & 13 \\
5 Oct. 2013                     & 2456571            & \FeII, [\FeII], and strong hydrogen lines in emission, & late B-type$^a$      & 14 \\
                                &                    & P-Cygni profiles                                       &                      & \\
                                &                    & \HeI\ very weakly in absorption                        &                      & \\
7 Oct. 2013                     & 2456573            & hydrogen in emission (weaker than in 10/2010),         & late A-type          & 12 \\
                                &                    & \CaII\ and \MgII\ in absorption,                       &                      & \\
                                &                    & \FeII\ in emission with P-Cygni profiles               &                      & \\
1/2 Nov. 2013                   & 2456598.5          & \FeII, [\FeII], and strong hydrogen lines in emission, & late B-type$^a$      & 14 \\
                                &                    & P-Cygni profiles                                       &                      & \\
                                &                    & \HeI\ very weakly in absorption                        &                      & \\
\hline
\end{tabular}
\tablefoot{\tablefoottext{a}{estimated spectral type based upon the description of the spectrum}}
\tablebib{(1) \citet{1953ApJ...118..353H}; (2) \citet{1975SvAL....1...30S};
(3) \citet{1975ApJ...200..426H}; (4) \citet{1978ApJ...219..445H};
(5) \citet{1985ApJ...290..542K}; (6) \citet{1988AA...203..306H};
(7) \citet{1996AA...314..131S}; (8) \citet{1995AJ....110.2715M};
(9) \citet{2012AA...541A.146C}; (10) this paper;
(11) \citet{2006AA...458..225V};
(12) \citet{2014ApJ...782L..21H}; (13) \citet{2013ATel.5403....1R};
(14) \citet{2013ATel.5538....1V}}
\end{table*}
}

\onllongtab{3}{
\begin{longtable}{l c c c c cl}
\caption{\label{Lightcurve_data_literature} Light curve data from the
  literature.}\\
\hline\hline
JD & $m_{pg}$ & B & B$_{err}$ & Reference \\
\hline
\endfirsthead
\caption{Continued.}\\
\hline\hline
JD & $m_{pg}$ & B & B$_{err}$ & Reference \\
\hline
\endhead
\hline
\endfoot
2414910       & 17.1  & 17.2   &       & \citet{1953ApJ...118..353H} \\
2416131       & 16.7  & 16.8   &       & ~~~~~~~~~~"  \\
2418969       & 17.3  & 17.4   &       & ~~~~~~~~~~"  \\
2420716       & 17.7  & 17.8   &       & ~~~~~~~~~~"  \\
2422503       & 17.3  & 17.4   &       & ~~~~~~~~~~"  \\
2422865       & 17.5  & 17.6   &       & ~~~~~~~~~~"  \\
2423248       & 17.4  & 17.5   &       & ~~~~~~~~~~"  \\
2423975       & 17.4  & 17.5   &       & ~~~~~~~~~~"  \\
2424333       & 17.4  & 17.5   &       & ~~~~~~~~~~"  \\
2424694       & 17.3  & 17.4   &       & ~~~~~~~~~~"  \\
2425056       & 17.5  & 17.6   &       & ~~~~~~~~~~"  \\
2425414       & 17.6  & 17.7   &       & ~~~~~~~~~~"  \\
2425776       & 17.2  & 17.3   &       & ~~~~~~~~~~"  \\
2426152       & 17.4  & 17.5   &       & ~~~~~~~~~~"  \\
2426528       & 17.5  & 17.6   &       & ~~~~~~~~~~"  \\
2426886       & 17.3  & 17.4   &       & ~~~~~~~~~~"  \\
2427620       & 17.4  & 17.5   &       & ~~~~~~~~~~"  \\
2427989       & 17.5  & 17.6   &       & ~~~~~~~~~~"  \\
2428340       & 17.3  & 17.4   &       & ~~~~~~~~~~"  \\
2428712       & 17.2  & 17.3   &       & ~~~~~~~~~~"  \\
2429077       & 17.2  & 17.3   &       & ~~~~~~~~~~"  \\
2429446       & 16.3  & 16.4   &       & ~~~~~~~~~~"  \\
2429808       & 16.6  & 16.7   &       & ~~~~~~~~~~"  \\
2431729       & 15.6  & 15.7   &       & ~~~~~~~~~~"  \\
2432039       & 15.3  & 15.4   &       & ~~~~~~~~~~"  \\
2432039       & 15.2  & 15.3   &       & ~~~~~~~~~~"  \\
2432102       & 15.4  & 15.5   &       & ~~~~~~~~~~"  \\
2432171       & 15.4  & 15.5   &       & ~~~~~~~~~~"  \\
2432379       & 15.5  & 15.6   &       & ~~~~~~~~~~"  \\
2432390       & 15.3  & 15.4   &       & ~~~~~~~~~~"  \\
2432437       & 15.5  & 15.6   &       & ~~~~~~~~~~"  \\
2432437       & 15.3  & 15.4   &       & ~~~~~~~~~~"  \\
2432583       & 15.3  & 15.4   &       & ~~~~~~~~~~"  \\
2432814       & 15.5  & 15.6   &       & ~~~~~~~~~~"  \\
2432887       & 15.3  & 15.4   &       & ~~~~~~~~~~"  \\
2433548       & 15.6  & 15.7   &       & ~~~~~~~~~~"  \\
2433617       & 15.9  & 16.0   &       & ~~~~~~~~~~"  \\
2433648       & 16.2  & 16.3   &       & ~~~~~~~~~~"  \\
2433975       & 16.5  & 16.6   &       & ~~~~~~~~~~"  \\
2433993       & 16.4  & 16.5   &       & ~~~~~~~~~~"  \\
2434015       & 16.6  & 16.7   &       & ~~~~~~~~~~"  \\
2434184       & 16.9  & 17.0   &       & ~~~~~~~~~~"  \\
2434227       & 17.0  & 17.1   &       & ~~~~~~~~~~"  \\
2434290       & 16.8  & 16.9   &       & ~~~~~~~~~~"  \\
2434379       & 16.8  & 16.9   &       & ~~~~~~~~~~"  \\
2434410       & 17.1  & 17.2   &       & ~~~~~~~~~~"  \\
2434443       & 16.9  & 17.0   &       & ~~~~~~~~~~"  \\
\hline
2428807       & 16.8  & 16.9   &       & \citet{1973PZ.....19....3S} \\
2428814       & 16.6  & 16.7   &       & ~~~~~~~~~~"  \\
2429230       & 16.6  & 16.7   &       & ~~~~~~~~~~"  \\
2429250       & 16.6  & 16.7   &       & ~~~~~~~~~~"  \\
2433891       & 16.6  & 16.7   &       & ~~~~~~~~~~"  \\
2433895       & 16.4  & 16.5   &       & ~~~~~~~~~~"  \\
2434330       & 16.8  & 16.9   &       & ~~~~~~~~~~"  \\
2435394       & 16.5  & 16.6   &       & ~~~~~~~~~~"  \\
2435431       & 16.7  & 16.8   &       & ~~~~~~~~~~"  \\
2436078       & 16.4  & 16.5   &       & ~~~~~~~~~~"  \\
2436821       & 16.8  & 16.9   &       & ~~~~~~~~~~"  \\
2437547       & 16.5  & 16.6   &       & ~~~~~~~~~~"  \\
2437911       & 16.7  & 16.8   &       & ~~~~~~~~~~"  \\
2437932       & 16.6  & 16.7   &       & \citet{1973PZ.....19....3S} \\
2437948       & 16.6  & 16.7   &       & ~~~~~~~~~~"  \\
2437963       & 16.6  & 16.7   &       & ~~~~~~~~~~"  \\
2438016       & 16.5  & 16.6   &       & ~~~~~~~~~~"  \\
2438292       & 16.4  & 16.5   &       & ~~~~~~~~~~"  \\
2439036       & 16.6  & 16.7   &       & ~~~~~~~~~~"  \\
2439418       & 16.20 & 16.30  &       & ~~~~~~~~~~"  \\
2439421       & 16.35 & 16.45  &       & ~~~~~~~~~~"  \\
2441217       & 16.7  & 16.8   &       & ~~~~~~~~~~"  \\
2441566       & 17.00 & 17.10  &       & ~~~~~~~~~~"  \\
2441570       & 16.80 & 16.90  &       & ~~~~~~~~~~"  \\
2441579       & 16.77 & 16.87  &       & ~~~~~~~~~~"  \\
\hline
2437212       & 17.0  & 17.1   &       & \citet{1973AA....22..453R} \\
2437237       & 17.1  & 17.2   &       & ~~~~~~~~~~"  \\
2437245       & 17.1  & 17.2   &       & ~~~~~~~~~~"  \\
2437289       & 17.1  & 17.2   &       & ~~~~~~~~~~"  \\
2437329       & 17.1  & 17.2   &       & ~~~~~~~~~~"  \\
2437373       & 17.1  & 17.2   &       & ~~~~~~~~~~"  \\
2437529       & 17.1  & 17.2   &       & ~~~~~~~~~~"  \\
2437570       & 17.0  & 17.1   &       & ~~~~~~~~~~"  \\
2437621       & 17.0  & 17.1   &       & ~~~~~~~~~~"  \\
2437916       & 16.9  & 17.0   &       & ~~~~~~~~~~"  \\
2437964       & 16.7  & 16.8   &       & ~~~~~~~~~~"  \\
2437978       & 16.7  & 16.8   &       & ~~~~~~~~~~"  \\
2438004       & 16.7  & 16.8   &       & ~~~~~~~~~~"  \\
2438031       & 16.7  & 16.8   &       & ~~~~~~~~~~"  \\
2438033       & 16.6  & 16.7   &       & ~~~~~~~~~~"  \\
2438618       & 16.5  & 16.6   &       & ~~~~~~~~~~"  \\
2439067       & 16.5  & 16.6   &       & ~~~~~~~~~~"  \\
2439436       & 16.5  & 16.6   &       & ~~~~~~~~~~"  \\
2439523       & 16.7  & 16.8   &       & ~~~~~~~~~~"  \\
2439523       & 16.6  & 16.7   &       & ~~~~~~~~~~"  \\
2440145       & 16.6  & 16.7   &       & ~~~~~~~~~~"  \\
2440178       & 16.6  & 16.7   &       & ~~~~~~~~~~"  \\
2440196       & 16.4  & 16.5   &       & ~~~~~~~~~~"  \\
2440200       & 16.4  & 16.5   &       & ~~~~~~~~~~"  \\
2440203       & 16.6  & 16.7   &       & ~~~~~~~~~~"  \\
2440214       & 16.4  & 16.5   &       & ~~~~~~~~~~"  \\
2440251       & 16.4  & 16.5   &       & ~~~~~~~~~~"  \\
2440506       & 16.8  & 16.9   &       & ~~~~~~~~~~"  \\
2440539       & 16.7  & 16.8   &       & ~~~~~~~~~~"  \\
2440550       & 16.8  & 16.9   &       & ~~~~~~~~~~"  \\
2440576       & 16.8  & 16.9   &       & ~~~~~~~~~~"  \\
2440831       & 16.9  & 17.0   &       & ~~~~~~~~~~"  \\
2440864       & 16.9  & 17.0   &       & ~~~~~~~~~~"  \\
2440889       & 16.9  & 17.0   &       & ~~~~~~~~~~"  \\
2440908       & 16.8  & 16.9   &       & ~~~~~~~~~~"  \\
2440937       & 16.9  & 17.0   &       & ~~~~~~~~~~"  \\
2440955       & 16.9  & 17.0   &       & ~~~~~~~~~~"  \\
2441010       & 17.0  & 17.1   &       & ~~~~~~~~~~"  \\
2441189       & 16.9  & 17.0   &       & ~~~~~~~~~~"  \\
2441225       & 16.9  & 17.0   &       & ~~~~~~~~~~"  \\
2441265       & 16.9  & 17.0   &       & ~~~~~~~~~~"  \\
2441269       & 16.9  & 17.0   &       & ~~~~~~~~~~"  \\
2441309       & 16.9  & 17.0   &       & ~~~~~~~~~~"  \\
2441324       & 16.9  & 17.0   &       & ~~~~~~~~~~"  \\
2441335       & 17.0  & 17.1   &       & ~~~~~~~~~~"  \\
2441536       & 17.0  & 17.1   &       & ~~~~~~~~~~"  \\
\hline
2439090.51    & 16.6  & 16.7 & $\approx$ 0.5 & \citet{1988IBVS.3193....1L} \\
2439498.38    & 16.3  & 16.4 & ~~"   & ~~~~~~~~~~"  \\
2439529.30    & 16.5  & 16.6 & ~~"   & ~~~~~~~~~~"  \\
2439711.55    & 16.1  & 16.2 & $\approx$ 0.5 & \citet{1988IBVS.3193....1L} \\
2439766.41    & 16.4  & 16.5 & ~~"   & ~~~~~~~~~~"  \\
2439796.44    & 16.5  & 16.6 & ~~"   & ~~~~~~~~~~"  \\
2439827.53    & 16.7  & 16.8 & ~~"   & ~~~~~~~~~~"  \\
2440073.56    & 16.4  & 16.5 & ~~"   & ~~~~~~~~~~"  \\
2440092.48    & 16.1  & 16.2 & ~~"   & ~~~~~~~~~~"  \\
2440144.49    & 16.6  & 16.7 & ~~"   & ~~~~~~~~~~"  \\
2440157.49    & 16.1  & 16.2 & ~~"   & ~~~~~~~~~~"  \\
2440183.42    & 16.5  & 16.6 & ~~"   & ~~~~~~~~~~"  \\
2440203.38    & 16.4  & 16.5 & ~~"   & ~~~~~~~~~~"  \\
2440230.24    & 16.4  & 16.5 & ~~"   & ~~~~~~~~~~"  \\
2440654.32    & 16.6  & 16.7 & ~~"   & ~~~~~~~~~~"  \\
2440798.54    & 16.6  & 16.7 & ~~"   & ~~~~~~~~~~"  \\
2440837.56    & 16.7  & 16.8 & ~~"   & ~~~~~~~~~~"  \\
2440916.47    & 16.7  & 16.8 & ~~"   & ~~~~~~~~~~"  \\
2441164.53    & 16.7  & 16.8 & ~~"   & ~~~~~~~~~~"  \\
2441183.50    & 16.7  & 16.8 & ~~"   & ~~~~~~~~~~"  \\
2441213.45    & 16.6  & 16.7 & ~~"   & ~~~~~~~~~~"  \\
2441518.55    & 16.4  & 16.5 & ~~"   & ~~~~~~~~~~"  \\
2441520.51    & 16.6  & 16.7 & ~~"   & ~~~~~~~~~~"  \\
2441625.46    & 16.6  & 16.7 & ~~"   & ~~~~~~~~~~"  \\
2441679.26    & 16.6  & 16.7 & ~~"   & ~~~~~~~~~~"  \\
2441687.38    & 16.5  & 16.6 & ~~"   & ~~~~~~~~~~"  \\
2441689.28    & 16.6  & 16.7 & ~~"   & ~~~~~~~~~~"  \\
2441714.34    & 16.5  & 16.6 & ~~"   & ~~~~~~~~~~"  \\
2441903.53    & 16.7  & 16.8 & ~~"   & ~~~~~~~~~~"  \\
2441921.55    & 16.7  & 16.8 & ~~"   & ~~~~~~~~~~"  \\
2442008.55    & 16.7  & 16.8 & ~~"   & ~~~~~~~~~~"  \\
2442066.40    & 16.7  & 16.8 & ~~"   & ~~~~~~~~~~"  \\
2442095.30    & 16.7  & 16.8 & ~~"   & ~~~~~~~~~~"  \\
2442278.44    & 16.7  & 16.8 & ~~"   & ~~~~~~~~~~"  \\
2442397.36    & 16.8  & 16.9 & ~~"   & ~~~~~~~~~~"  \\
2442473.31    & 16.8  & 16.9 & ~~"   & ~~~~~~~~~~"  \\
2442695.46    & 16.7  & 16.8 & ~~"   & ~~~~~~~~~~"  \\
2442725.47    & 16.8  & 16.9 & ~~"   & ~~~~~~~~~~"  \\
2442754.44    & 16.7  & 16.8 & ~~"   & ~~~~~~~~~~"  \\
2442756.50    & 16.8  & 16.9 & ~~"   & ~~~~~~~~~~"  \\
2443013.52    & 16.7  & 16.8 & ~~"   & ~~~~~~~~~~"  \\
2443072.49    & 16.7  & 16.8 & ~~"   & ~~~~~~~~~~"  \\
2443191.30    & 16.7  & 16.8 & ~~"   & ~~~~~~~~~~"  \\
2443344.57    & 16.6  & 16.7 & ~~"   & ~~~~~~~~~~"  \\
2443399.43    & 16.8  & 16.9 & ~~"   & ~~~~~~~~~~"  \\
2443430.50    & 16.6  & 16.7 & ~~"   & ~~~~~~~~~~"  \\
2443464.50    & 16.7  & 16.8 & ~~"   & ~~~~~~~~~~"  \\
2443489.29    & 16.7  & 16.8 & ~~"   & ~~~~~~~~~~"  \\
2443720.55    & 16.7  & 16.8 & ~~"   & ~~~~~~~~~~"  \\
2443756.43    & 16.7  & 16.8 & ~~"   & ~~~~~~~~~~"  \\
2443757.56    & 16.8  & 16.9 & ~~"   & ~~~~~~~~~~"  \\
2443787.54    & 16.8  & 16.9 & ~~"   & ~~~~~~~~~~"  \\
2443809.56    & 16.6  & 16.7 & ~~"   & ~~~~~~~~~~"  \\
2443815.28    & 16.6  & 16.7 & ~~"   & ~~~~~~~~~~"  \\
2444136.58    & 16.6  & 16.7 & ~~"   & ~~~~~~~~~~"  \\
2444167.47    & 16.6  & 16.7 & ~~"   & ~~~~~~~~~~"  \\
2444256.30    & 16.7  & 16.8 & ~~"   & ~~~~~~~~~~"  \\
2444554.47    & 16.6  & 16.7 & ~~"   & ~~~~~~~~~~"  \\
2444912.59    & 16.8  & 16.9 & ~~"   & ~~~~~~~~~~"  \\
2444989.29    & 16.6  & 16.7 & ~~"   & ~~~~~~~~~~"  \\
2445018.31    & 16.7  & 16.8 & ~~"   & ~~~~~~~~~~"  \\
2445197.53    & 16.8  & 16.9 & ~~"   & ~~~~~~~~~~"  \\
2445230.43    & 16.8  & 16.9 & ~~"   & ~~~~~~~~~~"  \\
2445261.46    & 16.7  & 16.8 & ~~"   & ~~~~~~~~~~"  \\
2445347.26    & 16.6  & 16.7 & $\approx$ 0.5 & \citet{1988IBVS.3193....1L} \\
2445593.50    & 16.0  & 16.1 & ~~"   & ~~~~~~~~~~"  \\
2445615.46    & 16.0  & 16.1 & ~~"   & ~~~~~~~~~~"  \\
2445647.49    & 15.6  & 15.7 & ~~"   & ~~~~~~~~~~"  \\
2445940.56    & 15.5  & 15.6 & ~~"   & ~~~~~~~~~~"  \\
2446026.43    & 15.4  & 15.5 & ~~"   & ~~~~~~~~~~"  \\
2446030.28    & 15.3  & 15.4 & ~~"   & ~~~~~~~~~~"  \\
2446321.43    & 15.4  & 15.5 & ~~"   & ~~~~~~~~~~"  \\
2446355.55    & 15.6  & 15.7 & ~~"   & ~~~~~~~~~~"  \\
2446441.35    & 15.3  & 15.4 & ~~"   & ~~~~~~~~~~"  \\
2446468.36    & 15.3  & 15.4 & ~~"   & ~~~~~~~~~~"  \\
2446677.57    & 15.4  & 15.5 & ~~"   & ~~~~~~~~~~"  \\
2446706.41    & 15.5  & 15.6 & ~~"   & ~~~~~~~~~~"  \\
2446738.50    & 15.7  & 15.8 & ~~"   & ~~~~~~~~~~"  \\
2446763.46    & 15.6  & 15.7 & ~~"   & ~~~~~~~~~~"  \\
2447060.48    & 15.5  & 15.6 & ~~"   & ~~~~~~~~~~"  \\
\hline
2439502.71    &       & 17.77  & $\approx$ 0.7 & \citet{1975ApJS...29..303V} \\
2439679.95    &       & 17.58  & ~~"   & ~~~~~~~~~~"  \\
2440209.73    &       & 17.38  & ~~"   & ~~~~~~~~~~"  \\
2440212.62    &       & 17.87  & ~~"   & ~~~~~~~~~~"  \\
2440212.68    &       & 17.40  & ~~"   & ~~~~~~~~~~"  \\
2440476.79    &       & 17.76  & ~~"   & ~~~~~~~~~~"  \\
2440476.86    &       & 17.60  & ~~"   & ~~~~~~~~~~"  \\
2440479.76    &       & 17.65  & ~~"   & ~~~~~~~~~~"  \\
2440858.75    &       & 18.10  & ~~"   & ~~~~~~~~~~"  \\
2440863.69    &       & 18.10  & ~~"   & ~~~~~~~~~~"  \\
2441245.86    &       & 17.95  & ~~"   & ~~~~~~~~~~"  \\
2441508.94    &       & 17.89  & ~~"   & ~~~~~~~~~~"  \\
2441512.92    &       & 16.92  & ~~"   & ~~~~~~~~~~"  \\
2441513.93    &       & 17.44  & ~~"   & ~~~~~~~~~~"  \\
2441514.85    &       & 17.52  & ~~"   & ~~~~~~~~~~"  \\
2441981.84    &       & 17.68  & ~~"   & ~~~~~~~~~~"  \\
2442363.72    &       & 17.68  & ~~"   & ~~~~~~~~~~"  \\
2442364.60    &       & 17.63  & ~~"   & ~~~~~~~~~~"  \\
2442365.63    &       & 17.40  & ~~"   & ~~~~~~~~~~"  \\
2442366.74    &       & 17.04  & ~~"   & ~~~~~~~~~~"  \\
\hline
2441572       & 16.71 & 16.81 &       & \citet{1990SvA....34..364S} \\
2441926       & 16.74 & 16.84 &       & ~~~~~~~~~~"  \\
2441956       & 16.78 & 16.88 &       & ~~~~~~~~~~"  \\
2441975       & 16.77 & 16.87 &       & ~~~~~~~~~~"  \\
2442316       & 16.77 & 16.87 &       & ~~~~~~~~~~"  \\
2442333       & 16.63 & 16.73 &       & ~~~~~~~~~~"  \\
2442364       & 17.32 & 17.42 &       & ~~~~~~~~~~"  \\
2442639       & 17.36 & 17.46 &       & ~~~~~~~~~~"  \\
2442659       & 17.30 & 17.40 &       & ~~~~~~~~~~"  \\
2442692       & 17.08 & 17.18 &       & ~~~~~~~~~~"  \\
2442721       & 17.06 & 17.16 &       & ~~~~~~~~~~"  \\
2443015       & 17.22 & 17.32 &       & ~~~~~~~~~~"  \\
2443400       & 17.26 & 17.36 &       & ~~~~~~~~~~"  \\
2443499       & 17.50 & 17.60 &       & ~~~~~~~~~~"  \\
2443782       & 17.32 & 17.42 &       & ~~~~~~~~~~"  \\
2444108       & 16.82 & 16.92 &       & ~~~~~~~~~~"  \\
2444140       & 16.92 & 17.02 &       & ~~~~~~~~~~"  \\
2444165       & 16.84 & 16.94 &       & ~~~~~~~~~~"  \\
2444461       & 16.72 & 16.82 &       & ~~~~~~~~~~"  \\
2444493       & 16.85 & 16.95 &       & ~~~~~~~~~~"  \\
2444828       & 16.77 & 16.87 &       & ~~~~~~~~~~"  \\
2444853       & 17.20 & 17.30 &       & ~~~~~~~~~~"  \\
2444880       & 17.12 & 17.22 &       & ~~~~~~~~~~"  \\
2445203       & 17.18 & 17.28 &       & ~~~~~~~~~~"  \\
2445234       & 16.94 & 17.04 &       & ~~~~~~~~~~"  \\
2445263       & 16.91 & 17.01 &       & \citet{1990SvA....34..364S} \\
2445286       & 16.84 & 16.94 &       & ~~~~~~~~~~"  \\
2445559       & 16.17 & 16.27 &       & ~~~~~~~~~~"  \\
2445588       & 15.83 & 15.93 &       & ~~~~~~~~~~"  \\
2445615       & 15.87 & 15.97 &       & ~~~~~~~~~~"  \\
2445703       & 15.72 & 15.82 &       & ~~~~~~~~~~"  \\
2445920       & 15.52 & 15.62 &       & ~~~~~~~~~~"  \\
2445944       & 15.50 & 15.60 &       & ~~~~~~~~~~"  \\
2445971       & 15.48 & 15.58 &       & ~~~~~~~~~~"  \\
2445994       & 15.50 & 15.60 &       & ~~~~~~~~~~"  \\
2446028       & 15.54 & 15.64 &       & ~~~~~~~~~~"  \\
2446298       & 15.60 & 15.70 &       & ~~~~~~~~~~"  \\
2446326       & 15.58 & 15.68 &       & ~~~~~~~~~~"  \\
2446358       & 15.48 & 15.58 &       & ~~~~~~~~~~"  \\
2446654       & 15.57 & 15.67 &       & ~~~~~~~~~~"  \\
2446684       & 15.56 & 15.66 &       & ~~~~~~~~~~"  \\
2446709       & 15.51 & 15.61 &       & ~~~~~~~~~~"  \\
2446764       & 15.50 & 15.60 &       & ~~~~~~~~~~"  \\
2447034       & 15.34 & 15.44 &       & ~~~~~~~~~~"  \\
2447062       & 16.05 & 16.15 &       & ~~~~~~~~~~"  \\
2447092       & 16.06 & 16.16 &       & ~~~~~~~~~~"  \\
2447124       & 16.00 & 16.10 &       & ~~~~~~~~~~"  \\
2447148       & 16.12 & 16.22 &       & ~~~~~~~~~~"  \\
2447419       & 16.33 & 16.43 &       & ~~~~~~~~~~"  \\
2447449       & 16.45 & 16.55 &       & ~~~~~~~~~~"  \\
\hline
2441900       &       & 16.81 &       & \citet{2004CoAst.145...28Z} \\
2441930       &       & 16.60 &       & ~~~~~~~~~~"  \\
2441950       &       & 16.69 &       & ~~~~~~~~~~"  \\
2442230       &       & 17.11 &       & ~~~~~~~~~~"  \\
2442280       &       & 16.87 &       & ~~~~~~~~~~"  \\
2442310       &       & 17.00 &       & ~~~~~~~~~~"  \\
2442330       &       & 17.31 &       & ~~~~~~~~~~"  \\
2442370       &       & 16.81 &       & ~~~~~~~~~~"  \\
2442380       &       & 17.23 &       & ~~~~~~~~~~"  \\
2442660       &       & 17.10 &       & ~~~~~~~~~~"  \\
2442680       &       & 17.50 &       & ~~~~~~~~~~"  \\
2442690       &       & 16.74 &       & ~~~~~~~~~~"  \\
2442750       &       & 17.02 &       & ~~~~~~~~~~"  \\
2442750       &       & 17.39 &       & ~~~~~~~~~~"  \\
2442750       &       & 16.89 &       & ~~~~~~~~~~"  \\
2443020       &       & 17.15 &       & ~~~~~~~~~~"  \\
2443370       &       & 16.71 &       & ~~~~~~~~~~"  \\
2443380       &       & 17.19 &       & ~~~~~~~~~~"  \\
2443380       &       & 17.01 &       & ~~~~~~~~~~"  \\
2443740       &       & 16.86 &       & ~~~~~~~~~~"  \\
2443760       &       & 17.17 &       & ~~~~~~~~~~"  \\
2444080       &       & 17.39 &       & ~~~~~~~~~~"  \\
2444110       &       & 17.20 &       & ~~~~~~~~~~"  \\
2444130       &       & 16.84 &       & ~~~~~~~~~~"  \\
2444470       &       & 17.22 &       & ~~~~~~~~~~"  \\
2444480       &       & 16.82 &       & ~~~~~~~~~~"  \\
2444500       &       & 17.36 &       & ~~~~~~~~~~"  \\
2444820       &       & 16.99 &       & ~~~~~~~~~~"  \\
2444880       &       & 17.20 &       & ~~~~~~~~~~"  \\
2445220       &       & 17.09 &       & ~~~~~~~~~~"  \\
2445250       &       & 16.90 &       & ~~~~~~~~~~"  \\
2445250       &       & 16.72 &       & ~~~~~~~~~~"  \\
2445260       &       & 16.56 &       & ~~~~~~~~~~"  \\
2445470       &       & 15.91 &       & ~~~~~~~~~~"  \\
2445530       &       & 15.77 &       & ~~~~~~~~~~"  \\
2445540       &       & 15.64 &       & ~~~~~~~~~~"  \\
2445610       &       & 16.05 &       & \citet{2004CoAst.145...28Z} \\
2445730       &       & 15.91 &       & ~~~~~~~~~~"  \\
2445940       &       & 15.75 &       & ~~~~~~~~~~"  \\
2445990       &       & 15.49 &       & ~~~~~~~~~~"  \\
2446310       &       & 15.84 &       & ~~~~~~~~~~"  \\
2446360       &       & 15.50 &       & ~~~~~~~~~~"  \\
2446690       &       & 15.67 &       & ~~~~~~~~~~"  \\
2446720       &       & 15.59 &       & ~~~~~~~~~~"  \\
2447080       &       & 16.09 &       & ~~~~~~~~~~"  \\
2447080       &       & 16.29 &       & ~~~~~~~~~~"  \\
2447350       &       & 16.51 &       & ~~~~~~~~~~"  \\
2447450       &       & 16.08 &       & ~~~~~~~~~~"  \\
2447470       &       & 16.30 &       & ~~~~~~~~~~"  \\
2447670       &       & 16.75 &       & ~~~~~~~~~~"  \\
2447800       &       & 16.87 &       & ~~~~~~~~~~"  \\
2447880       &       & 16.40 &       & ~~~~~~~~~~"  \\
2448140       &       & 16.85 &       & ~~~~~~~~~~"  \\
2448180       &       & 16.71 &       & ~~~~~~~~~~"  \\
2448250       &       & 16.31 &       & ~~~~~~~~~~"  \\
2448510       &       & 16.39 &       & ~~~~~~~~~~"  \\
2448520       &       & 16.62 &       & ~~~~~~~~~~"  \\
2448570       &       & 16.00 &       & ~~~~~~~~~~"  \\
2448590       &       & 16.08 &       & ~~~~~~~~~~"  \\
2448640       &       & 16.51 &       & ~~~~~~~~~~"  \\
2448850       &       & 15.80 &       & ~~~~~~~~~~"  \\
2448900       &       & 16.46 &       & ~~~~~~~~~~"  \\
2448900       &       & 16.71 &       & ~~~~~~~~~~"  \\
2448940       &       & 16.01 &       & ~~~~~~~~~~"  \\
2449190       &       & 16.52 &       & ~~~~~~~~~~"  \\
2449200       &       & 16.19 &       & ~~~~~~~~~~"  \\
2449210       &       & 16.68 &       & ~~~~~~~~~~"  \\
2449250       &       & 16.80 &       & ~~~~~~~~~~"  \\
2449580       &       & 16.72 &       & ~~~~~~~~~~"  \\
2449630       &       & 16.31 &       & ~~~~~~~~~~"  \\
2449680       &       & 16.20 &       & ~~~~~~~~~~"  \\
2449690       &       & 16.44 &       & ~~~~~~~~~~"  \\
2449980       &       & 16.77 &       & ~~~~~~~~~~"  \\
2450040       &       & 16.46 &       & ~~~~~~~~~~"  \\
2450090       &       & 16.62 &       & ~~~~~~~~~~"  \\
2450290       &       & 16.15 &       & ~~~~~~~~~~"  \\
2450340       &       & 16.47 &       & ~~~~~~~~~~"  \\
2450370       &       & 15.91 &       & ~~~~~~~~~~"  \\
2450380       &       & 16.02 &       & ~~~~~~~~~~"  \\
2450440       &       & 16.18 &       & ~~~~~~~~~~"  \\
2450610       &       & 16.64 &       & ~~~~~~~~~~"  \\
2450670       &       & 16.81 &       & ~~~~~~~~~~"  \\
2450670       &       & 16.31 &       & ~~~~~~~~~~"  \\
2451020       &       & 16.86 &       & ~~~~~~~~~~"  \\
2451090       &       & 16.42 &       & ~~~~~~~~~~"  \\
2451380       &       & 17.01 &       & ~~~~~~~~~~"  \\
2451430       &       & 16.59 &       & ~~~~~~~~~~"  \\
2451510       &       & 16.87 &       & ~~~~~~~~~~"  \\
2451530       &       & 16.71 &       & ~~~~~~~~~~"  \\
2451750       &       & 16.45 &       & ~~~~~~~~~~"  \\
2451790       &       & 16.83 &       & ~~~~~~~~~~"  \\
2451800       &       & 16.62 &       & ~~~~~~~~~~"  \\
2451820       &       & 16.31 &       & ~~~~~~~~~~"  \\
2452140       &       & 16.10 &       & ~~~~~~~~~~"  \\
2452200       &       & 16.56 &       & ~~~~~~~~~~"  \\
2452250       &       & 16.40 &       & ~~~~~~~~~~"  \\
2452480       &       & 16.20 &       & ~~~~~~~~~~"  \\
2452520       &       & 15.97 &       & \citet{2004CoAst.145...28Z} \\
2452550       &       & 15.76 &       & ~~~~~~~~~~"  \\
2452600       &       & 15.59 &       & ~~~~~~~~~~"  \\
2452610       &       & 15.45 &       & ~~~~~~~~~~"  \\
\hline
2443044       &       & 17.12  & 0.01 &\citet{1978ApJ...219..445H} \\
2443428       &       & 16.98  & 0.01 & \citet{1978ApJ...221L..73H} \\
2444559$\pm$15&       & 17.20  & 0.01 & \citet{1984ApJ...278..124H} \\
2444954$\pm$15&       & 17.33  & 0.01 & ~~~~~~~~~~"  \\
2445197$\pm$15&       & 17.24  & 0.01 & ~~~~~~~~~~"  \\
2445258$\pm$15&       & 16.57  & 0.03 & ~~~~~~~~~~"  \\
2446750$\pm$15&       & 16.12  & 0.03 & \citet{1988AA...203..306H} \\
2447054$\pm$15&       & 16.12  & 0.08 & ~~~~~~~~~~"  \\
\hline
2445286       &       & 17.47  & 0.15  & \citet{1999AA...349..796K} \\
2445295       &       & 17.43  & 0.15  & ~~~~~~~~~~"  \\
2445296       &       & 17.32  & 0.23  & ~~~~~~~~~~"  \\
2445297       &       & 17.39  & 0.25  & ~~~~~~~~~~"  \\
2445588       &       & 15.74  & 0.23  & ~~~~~~~~~~"  \\
2445588       &       & 16.04  & 0.31  & ~~~~~~~~~~"  \\
2445590       &       & 15.94  & 0.20  & ~~~~~~~~~~"  \\
2445591       &       & 15.69  & 0.16  & ~~~~~~~~~~"  \\
2445623       &       & 15.72  & 0.23  & ~~~~~~~~~~"  \\
2445625       &       & 15.90  & 0.16  & ~~~~~~~~~~"  \\
2445702       &       & 15.60  & 0.08  & ~~~~~~~~~~"  \\
2445929       &       & 15.55  & 0.08  & ~~~~~~~~~~"  \\
2445968       &       & 15.41  & 0.09  & ~~~~~~~~~~"  \\
2446435       &       & 15.47  & 0.08  & ~~~~~~~~~~"  \\
2446707       &       & 15.64  & 0.16  & ~~~~~~~~~~"  \\
2446707       &       & 15.46  & 0.16  & ~~~~~~~~~~"  \\
2446708       &       & 15.51  & 0.07  & ~~~~~~~~~~"  \\
2446708       &       & 15.40  & 0.10  & ~~~~~~~~~~"  \\
2446709       &       & 15.58  & 0.08  & ~~~~~~~~~~"  \\
2446738       &       & 15.66  & 0.11  & ~~~~~~~~~~"  \\
2446738       &       & 15.40  & 0.17  & ~~~~~~~~~~"  \\
2448177       &       & 16.36  & 0.18  & ~~~~~~~~~~"  \\
2448180       &       & 16.22  & 0.17  & ~~~~~~~~~~"  \\
2448180       &       & 16.22  & 0.13  & ~~~~~~~~~~"  \\
2448180       &       & 16.42  & 0.12  & ~~~~~~~~~~"  \\
\hline
2446703$\pm$2 &       & 16.43  &       & \citet{1995AJ....110.2715M} \\
\hline
2447416$\pm$33&       & 16.51  & 0.02  & \citet{1991AJ....101.1663W} \\
\hline
2448867       &       & 16.29  & 0.02--0.10 & \citet{1996AA...314..131S} \\
2448871       &       & 16.27  & ~~~"       & ~~~~~~~~~~"  \\
2448892       &       & 16.33  & ~~~"       & ~~~~~~~~~~"  \\
2448973       &       & 16.59  & ~~~"       & ~~~~~~~~~~"  \\
2448979       &       & 16.48  & ~~~"       & ~~~~~~~~~~"  \\
2449008       &       & 16.39  & ~~~"       & ~~~~~~~~~~"  \\
\hline
2451452.7982  &       & 17.179 & 0.001 & \citet{2001AJ....122.2477M} \\
2451452.9166  &       & 17.188 & 0.001 & ~~~~~~~~~~"  \\
2451454.6968  &       & 17.201 & 0.001 & ~~~~~~~~~~"  \\
2451454.7299  &       & 17.198 & 0.001 & ~~~~~~~~~~"  \\
2451454.8004  &       & 17.189 & 0.002 & ~~~~~~~~~~"  \\
2451454.8698  &       & 17.194 & 0.002 & ~~~~~~~~~~"  \\
2451454.8784  &       & 17.197 & 0.002 & ~~~~~~~~~~"  \\
2451454.9880  &       & 17.208 & 0.001 & ~~~~~~~~~~"  \\
2451455.7628  &       & 17.188 & 0.002 & ~~~~~~~~~~"  \\
2451455.8708  &       & 17.200 & 0.002 & ~~~~~~~~~~"  \\
2451456.6809  &       & 17.218 & 0.002 & ~~~~~~~~~~"  \\
2451456.7014  &       & 17.208 & 0.001 & ~~~~~~~~~~"  \\
2451456.7246  &       & 17.214 & 0.001 & ~~~~~~~~~~"  \\
2451456.7484  &       & 17.228 & 0.003 & ~~~~~~~~~~"  \\
2451456.7666  &       & 17.231 & 0.003 & ~~~~~~~~~~"  \\
2451456.7939  &       & 17.224 & 0.002 & ~~~~~~~~~~"  \\
2451456.8207  &       & 17.213 & 0.002 & \citet{2001AJ....122.2477M} \\
2451456.8403  &       & 17.202 & 0.001 & ~~~~~~~~~~"  \\
2451457.6993  &       & 17.187 & 0.002 & ~~~~~~~~~~"  \\
2451457.8141  &       & 17.209 & 0.002 & ~~~~~~~~~~"  \\
2451457.8759  &       & 17.202 & 0.001 & ~~~~~~~~~~"  \\
2451457.9503  &       & 17.215 & 0.002 & ~~~~~~~~~~"  \\
2451484.6991  &       & 17.181 & 0.001 & ~~~~~~~~~~"  \\
2451484.7858  &       & 17.177 & 0.001 & ~~~~~~~~~~"  \\
2451484.8954  &       & 17.166 & 0.001 & ~~~~~~~~~~"  \\
2451486.6341  &       & 17.181 & 0.002 & ~~~~~~~~~~"  \\
2451486.7428  &       & 17.185 & 0.001 & ~~~~~~~~~~"  \\
2451488.6521  &       & 17.160 & 0.001 & ~~~~~~~~~~"  \\
2451488.7601  &       & 17.185 & 0.001 & ~~~~~~~~~~"  \\
2451488.7806  &       & 17.178 & 0.001 & ~~~~~~~~~~"  \\
\hline
2453347       &       & 16.5   & 0.2   & \citet{2006AA...458..225V} \\
2453641       &       & 16.6   & 0.1   & ~~~~~~~~~~"  \\
\hline
2456536.78    &       & 15.87  &       & \citet{2013ATel..5362....1H} \\
\hline
2456598       &       & 15.9   &       & \citet{2013ATel.5538....1V}\\
\end{longtable}
}

\onllongtab{4}{
\begin{longtable}{l c c c c l}
\caption{\label{Lightcurve_data} Light curve data.}\\
\hline\hline
JD & B & B$_{err}$ & V & V$_{err}$ & Observatory \\
\hline
\endfirsthead
\caption{Continued.}\\
\hline\hline
JD & B & B$_{err}$ & V & V$_{err}$ & Observatory \\
\hline
\endhead
\hline
\endfoot
2421573       & 17.0 & 0.4 & & & HDAP \\
2421573       & 17.3 & 0.4 & & & ~"  \\
2417797       & 17.5 & 0.1 & & & ~"  \\
2417800       & 17.2 & 0.2 & & & ~"  \\
2418149       & 17.3 & 0.2 & & & ~"  \\
2421580       & 17.5 & 0.1 & & & ~"  \\
2421580       & 17.4 & 0.1 & & & ~"  \\
2421624       & 17.5 & 0.1 & & & ~"  \\
2421624       & 17.5 &           & & & ~"  \\
2422552       & 16.9 & 0.4 & & & ~"  \\
2422552       & 17.3 & 0.1 & & & ~"  \\
2423342       & 17.1 &           & & & ~"  \\
2423353       & 17.1 & 0.1 & & & ~"  \\
\hline
2438266.55    &    &     & 16.3 & 0.2 & TLS plates\\
2438268.49    & 16.7  & 0.2 & 16.4 & 0.2 & ~"  \\
2438286.52    & 15.8  & 0.2 & 16.1 & 0.2 & ~"  \\
2438287.49    & 15.9  & 0.4 & 15.0 & 0.3 & ~"  \\
2438289.51    & 16.3  & 0.3 & 16.2 & 0.1 & ~"  \\
2440500.46    &    &     & 16.1 & 0.1 & ~"  \\
2440501.54    &    &     & 16.4 & 0.1 & ~"  \\
2440503.52    &    &     & 16.6 & 0.1 & ~"  \\
2440504.57    &    &     & 16.4 &           & ~"  \\
2440506.43    &    &     & 16.0 & 0.2 & ~"  \\
2440508.50    & 16.8  & 0.2 & & & ~"  \\
2442805.29    & 16.9  & 0.2 & & & ~"  \\
2443816.33    & 16.8  & 0.2 & & & ~"  \\
2444194.44    & 17.2  & 0.2 & & & ~"  \\
2444222.40    & 17.2  & 0.2 & & & ~"  \\
2444253.31    & 16.1  & 0.4 & & & ~"  \\
2444489.56    & 16.9  & 0.2 & & & ~"  \\
2444490.54    & 16.8  & 0.3 & & & ~"  \\
2444491.57    & 17.2  & 0.2 & & & ~"  \\
2444523.37    & 16.9  & 0.2 & & & ~"  \\
2444500.58    & 16.8  & 0.3 & & & ~"  \\
2444544.36    & 17.4  & 0.2 & & & ~"  \\
2444545.34    & 17.3  & 0.3 & & & ~"  \\
2444912.39    & 16.8  & 0.3 & & & ~"  \\
2444912.42    & 16.5  & 0.3 & & & ~"  \\
2444912.46    & 16.5  & 0.3 & & & ~"  \\
2444912.48    & 16.7  & 0.3 & & & ~"  \\
2444933.41    & 17.1  & 0.2 & & & ~"  \\
2444933.44    & 17.1  & 0.2 & & & ~"  \\
2444933.46    & 16.9  & 0.2 & & & ~"  \\
2444967.34    & 17.0  & 0.2 & & & ~"  \\
2444967.35    & 17.2  & 0.2 & & & ~"  \\
2444967.37    & 17.1  & 0.2 & & & ~"  \\
2444967.38    & 17.0  & 0.2 & & & ~"  \\
2444967.39    & 16.9  & 0.2 & & & ~"  \\
2444988.36    & 16.8  & 0.2 & & & ~"  \\
2444988.37    & 17.0  & 0.2 & & & ~"  \\
2444988.38    & 16.9  & 0.2 & & & ~"  \\
2444989.32    & 16.9  & 0.3 & & & ~"  \\
2444989.34    & 17.0  & 0.3 & & & ~"  \\
2445020.28    & 17.0  & 0.2 & & & ~"  \\
2445370.25    & 16.3  & 0.2 & & & ~"  \\
2445370.27    & 16.3  & 0.2 & & & ~"  \\
2445370.30    & 16.3  & 0.2 & & & ~"  \\
2445372.30    & 16.3  & 0.2 & & & ~"  \\
2446320.58    & 15.7  & 0.3 & & & ~"  \\
2446321.62    & 16.5  & 0.2 & & & ~"  \\
2446650.56    & 16.0  & 0.2 & & & TLS plates \\
2446678.55    & 17.0  & 0.2 & & & ~"  \\
2446678.58    & 17.0  & 0.2 & & & ~"  \\
2446678.60    & 17.0  & 0.2 & & & ~"  \\
2446683.52    & 16.1  & 0.2 & & & ~"  \\
2446683.55    & 16.1  & 0.2 & & & ~"  \\
2446683.58    & 16.2  & 0.2 & & & ~"  \\
2446714.60    & 15.9  & 0.2 & & & ~"  \\
2446714.62    & 15.7  & 0.2 & & & ~"  \\
2446716.54    & 16.0  & 0.2 & & & ~"  \\
2446716.57    & 16.0  & 0.2 & & & ~"  \\
2446731.37    & 16.4  & 0.2 & & & ~"  \\
2446731.42    & 16.1  & 0.2 & & & ~"  \\
2446737.46    & 16.3  & 0.2 & & & ~"  \\
2446737.59    & 16.1  & 0.2 & & & ~"  \\
2446742.44    & 16.7  & 0.2 & & & ~"  \\
2446762.38    & 16.0  & 0.3 & & & ~"  \\
2446763.30    & 16.5  & 0.3 & & & ~"  \\
2446764.43    & 16.3  & 0.2 & & & ~"  \\
2446765.46    & 16.1  & 0.2 & & & ~"  \\
2447039.50    & 16.1  & 0.1 & & & ~"  \\
2447042.62    & 15.8  & 0.1 & & & ~"  \\
2447775.49    & 16.7  & 0.1 & & & ~"  \\
2447776.48    & 16.7  & 0.1 & & & ~"  \\
2447777.47    & 16.1  & 0.1 & & & ~"  \\
2447823.34    & 16.4  & 0.2 & & & ~"  \\
2447823.36    & 16.6  & 0.1 & & & ~"  \\
2447825.37    & 16.8  & 0.1 & & & ~"  \\
2447827.34    & 16.5  & 0.2 & & & ~"  \\
2448893.53    & 16.4  &     & & & ~"  \\
2448927.52    & 16.4  & 0.2 & & & ~"  \\
2448928.53    & 16.3  & 0.1 & & & ~"  \\
2448950.50    & 16.3  & 0.1 & & & ~"  \\
2449658.44    & 16.7  & 0.1 & & & ~"  \\
2450048.40    & 16.6  & 0.2 & & & ~"  \\
2450432.39    & 16.2  & 0.2 & & & ~"  \\
\hline
2450332.85    & 17.31  & 0.03 & & & DIRECT \\
2450334.90    & 16.99  & 0.34 & & & ~"    \\
2450365.80    & 17.25  & 0.01 & & & ~"    \\
2450688.97    & 17.84  & 0.01 & & & ~"    \\
2450690.96    & 17.79  & 0.01 & & & ~"    \\
2450691.93    & 17.68  & 0.01 & & & ~"    \\
2450694.87    & 17.62  & 0.01 & & & ~"    \\
2450694.99    & 17.75  & 0.01 & & & ~"    \\
2450695.93    & 17.62  & 0.01 & & & ~"    \\
2450695.94    & 17.03  & 0.05 & & & ~"    \\
2450696.92    & 17.72  & 0.02 & & & ~"    \\
2450696.99    & 17.81  & 0.01 & & & ~"    \\
2450730.93    & 17.80  & 0.01 & & & ~"    \\
2451044.85    &        &      & 17.18  & 0.17 & ~"    \\
2451051.95    & 17.80  & 0.04 & 17.00  & 0.18 & ~"    \\
2451055.86    &        &      & 17.96  & 0.07 & ~"    \\
2451056.99    &        &      & 17.95  & 0.33 & ~"    \\
2451057.88    &        &      & 17.26  & 0.17 & ~"    \\
2451074.93    &        &      & 17.91  & 0.18 & ~"    \\
2451101.89    &        &      & 18.37  & 0.16 & ~"    \\
2451102.81    &        &      & 18.15  & 0.18 & ~"    \\
2451103.82    &        &      & 18.18  & 0.12 & ~"    \\
2451104.85    &        &      & 16.05  & 0.15 & ~"    \\
2451105.96    &        &      & 16.12  & 0.14 & ~"    \\
2451106.70    &        &      & 16.88  & 0.15 & ~"    \\
2451138.75    & 18.04  & 0.01 & 17.97  & 0.18 & DIRECT \\
2451139.88    &        &      & 16.39  & 0.35 & ~"    \\
2451140.77    &        &      & 17.65  & 0.21 & ~"    \\
2451141.76    &        &      & 16.55  & 0.16 & ~"    \\
2451142.87    &        &      & 17.34  & 0.18 & ~"    \\
2451143.80    &        &      & 18.42  & 0.11 & ~"    \\
\hline
2451821.4     & 16.65 & 0.01 & 16.53 & 0.01 & NOAO \\
2452170.3     & 16.45 & 0.01 & 16.33 & 0.01 & ~"   \\
\hline
2452499.91    & 17.03  & 0.07 & 15.85 & 0.03 & WIYN  \\
2452499.95    & 16.22  & 0.19 & 15.93 & 0.09 & ~"    \\
2452517.79    & 16.79  & 0.13 & 15.78 & 0.06 & ~"    \\
2452517.80    & 16.64  & 0.09 & 15.74 & 0.04 & ~"    \\
2452531.74    & 17.12  & 0.15 & 15.82 & 0.07 & ~"    \\
2452531.78    & 16.75  & 0.05 & 15.93 & 0.09 & ~"    \\
2452557.79    & 16.03  & 0.16 & 15.73 & 0.08 & ~"    \\
2452557.80    & 16.82  & 0.16 & 15.98 & 0.13 & ~"    \\
2452589.80    & 16.03  & 0.17 & 15.83 & 0.08 & ~"    \\
2452589.80    & 16.51  & 0.12 & 15.85 & 0.10 & ~"    \\
2452646.63    & 17.09  & 0.13 & 15.87 & 0.08 & ~"    \\
2452646.64    & 16.76  & 0.08 & 15.80 & 0.06 & ~"    \\
2452662.60    & 18.05  & 0.70 & 15.83 & 0.07 & ~"    \\
2452662.61    & 16.32  & 0.09 & 15.75 & 0.06 & ~"    \\
2452901.76    & 17.02  & 0.21 & 15.76 & 0.03 & ~"    \\
2452901.77    & 16.61  & 0.23 & 15.99 & 0.11 & ~"    \\
2452908.80    & 16.09  & 0.12 & 15.85 & 0.05 & ~"    \\
2452908.81    & 16.33  & 0.08 & 15.87 & 0.06 & ~"    \\
2452932.80    & 16.48  & 0.02 & 15.54 & 0.01 & ~"    \\
2452932.80    & 16.52  & 0.01 & 15.58 & 0.02 & ~"    \\
2452958.81    & 16.10  & 0.10 & 15.95 & 0.08 & ~"    \\
2452958.82    & 16.58  & 0.06 & 15.80 & 0.06 & ~"    \\
2453000.65    & 16.80  & 0.06 & 15.73 & 0.03 & ~"    \\
2453000.65    &        &      & 15.75 & 0.03 & ~"    \\
2453567.95    & 16.93  & 0.16 & 16.24 & 0.06 & ~"    \\
2453567.95    & 16.87  & 0.14 & 16.29 & 0.09 & ~"    \\
\hline
2453764.3     & 16.80 & 0.01 &             &            & TLS CCD \\
2454387.5     & 17.02 & 0.02 & 16.94 & 0.01 & ~"   \\
2454472.3     & 17.17 & 0.01 & 17.08 & 0.01 & ~"   \\
2454742.3     & 17.15 & 0.02 & 17.07 & 0.01 & ~"   \\
2454823       & 17.20 & 0.02 &             &            & ~"   \\
2454852       & 17.19 & 0.02 & 17.10 & 0.02 & ~"   \\
2454891       & 17.25 & 0.04 & 17.20 & 0.02 & ~"   \\
2455094.5     & 17.56 & 0.01 &             &            & ~"   \\
2455124.4     & 17.56 & 0.02 &             &            & ~"   \\
2455125.6     & 17.54 & 0.02 & 17.48 & 0.01 & ~"   \\
2455126.6     & 17.53 & 0.02 & 17.47 & 0.01 & ~"   \\
2455248.3     & 17.53 & 0.02 & 17.51 & 0.01 & ~"   \\
2455621.0     & 16.40 & 0.02 & 16.21 & 0.02 & ~"   \\
2455806.1     & 16.13 & 0.01 & 15.81 & 0.01 & ~"   \\
2455834.5     & 15.98 & 0.01 & 15.79 & 0.01 & ~"   \\
2455835.5     & 16.07 & 0.01 & 15.79 & 0.01 & ~"   \\
2455838.4     & 16.05 & 0.07 & 15.84 & 0.04 & ~"   \\
\hline
2453943.6     & 17.03 & 0.01    & & & CAHA \\
2453983.6     & 16.84 & $<$0.01 & & & ~"   \\
2454031.3     & 16.55 & $<$0.01 & & & ~"   \\
\hline
2454376.4652  &        &      & 16.87 & & COoSAI \\
2454376.4689  &        &      & 16.88 & & ~"    \\
2454376.4719  &        &      & 16.85 & & ~"    \\
2454376.4748  &        &      & 16.91 & & ~"    \\
2454376.4777  &        &      & 16.86 & & ~"    \\
2454376.4806  &        &      & 16.88 & & ~"    \\
2454376.4864  &        &      & 16.88 & & ~"    \\
2454376.4893  &        &      & 16.91 & & COoSAI \\
2454376.4923  &        &      & 16.88 & & ~"    \\
2454376.4952  &        &      & 16.90 & & ~"    \\
2454376.4981  &        &      & 16.90 & & ~"    \\
2454376.5010  &        &      & 16.86 & & ~"    \\
2454376.5068  &        &      & 16.83 & & ~"    \\
2454376.5137  & 16.96  &      &       & & ~"    \\
2454376.5197  & 17.02  &      &       & & ~"    \\
2454385.4800  &        &      & 16.86 & & ~"    \\
2454385.4825  &        &      & 16.81 & & ~"    \\
2454385.4838  & 17.00  &      &       & & ~"    \\
2454389.4070  &        &      & 16.87 & & ~"    \\
2454389.4102  &        &      & 16.85 & & ~"    \\
2454389.4110  &        &      & 16.81 & & ~"    \\
2454389.4118  &        &      & 16.85 & & ~"    \\
2454389.4131  & 17.03  &      &       & & ~"    \\
2454389.4146  & 16.99  &      &       & & ~"    \\
2454389.4161  & 16.96  &      &       & & ~"    \\
2454391.4144  &        &      & 16.89 & & ~"    \\
2454391.4152  &        &      & 16.86 & & ~"    \\
2454391.4160  &        &      & 16.85 & & ~"    \\
2454391.4173  & 16.20  &      &       & & ~"    \\
2454391.4188  & 16.29  &      &       & & ~"    \\
2454391.4203  & 16.23  &      &       & & ~"    \\
2455141.4706  & 17.18  &      &       & & ~"    \\
2455141.4724  &        &      & 16.33 & & ~"    \\
2455144.3362  & 17.19  &      &       & & ~"    \\
2455144.3378  &        &      & 17.30 & & ~"    \\
2455144.3394  & 17.23  &      &       & & ~"    \\
2455144.3410  &        &      & 17.37 & & ~"    \\
2455144.3426  & 17.32  &      &       & & ~"    \\
2455144.3442  &        &      & 17.34 & & ~"    \\
2455144.3458  & 17.29  &      &       & & ~"    \\
2455144.3474  &        &      & 17.36 & & ~"    \\
2455144.3490  & 17.26  &      &       & & ~"    \\
2455144.3506  &        &      & 17.37 & & ~"    \\
2455144.3522  & 17.24  &      &       & & ~"    \\
2455144.3538  &        &      & 17.41 & & ~"    \\
2455144.3570  &        &      & 17.33 & & ~"    \\
2455144.3586  & 17.24  &      &       & & ~"    \\
2455144.3602  &        &      & 17.41 & & ~"    \\
2455144.3618  & 17.40  &      &       & & ~"    \\
2455144.3634  &        &      & 17.54 & & ~"    \\
\hline
2456139.89    &        &      & 15.64 & & John Martin (priv. com.) \\
2456209.65    &        &      & 15.85 & & ~~~~~~~~~~"  \\
2456274.61    &        &      & 15.86 & & ~~~~~~~~~~"  \\
\end{longtable}
}

\begin{figure*}
   \centering
   \includegraphics[width=\textwidth]{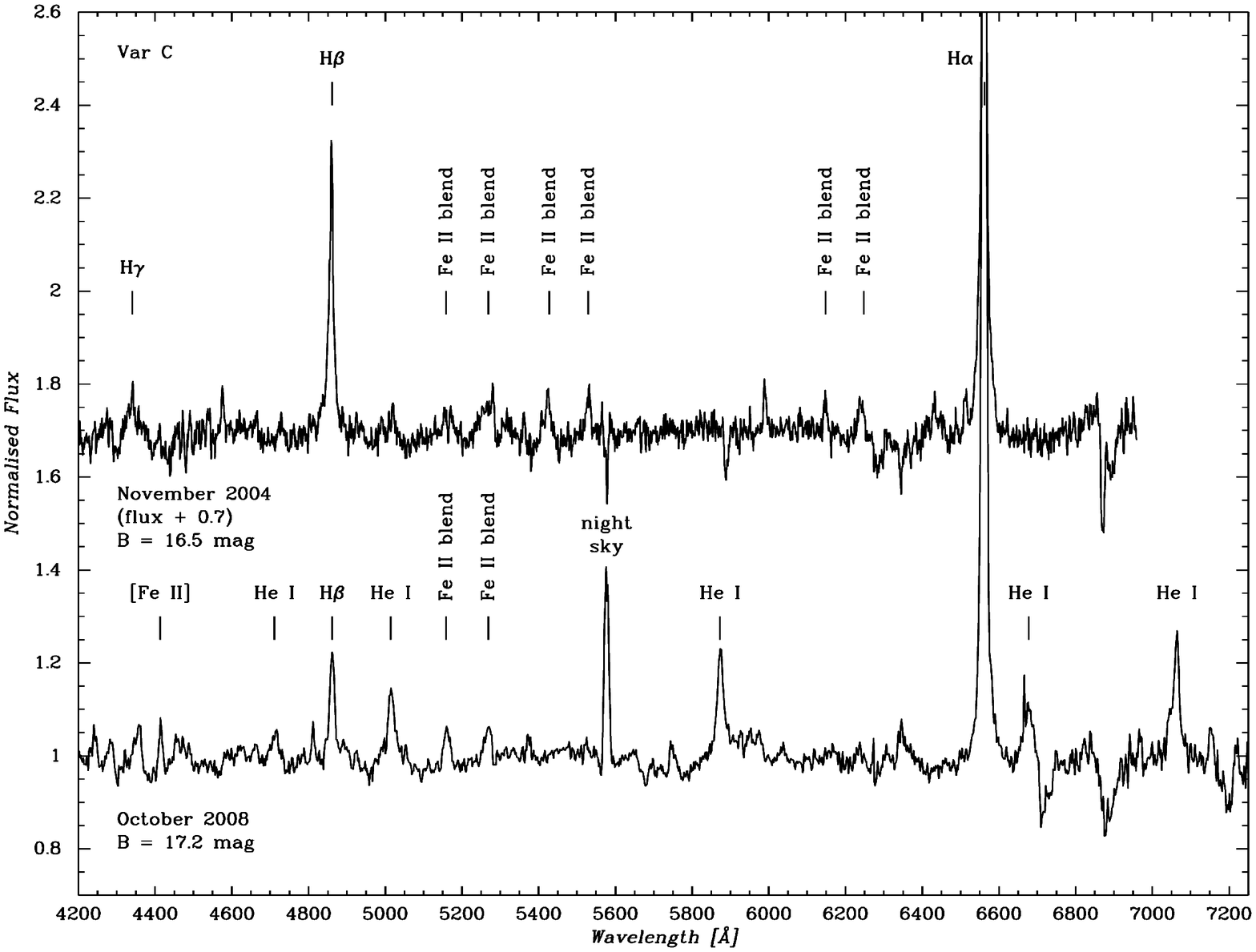}
   \caption{SAO RAS spectra of Var~C taken on 13 November 2004 (upper
     spectrum) and 22 October 2008 (lower spectrum).}
   \label{spec_varc}
\end{figure*}

A light curve of Var~C between 1899 and 2013 containing values from the
literature, photometry done on archival data, and our observations is presented
in Figure \ref{licu_VarC}. The corresponding data values are listed in Tables
\ref{Lightcurve_data_literature} and \ref{Lightcurve_data}.

The earliest four data points found in the literature spread over almost 20
years and vary over one magnitude. 
Low data coverage during that time prevented us from reconstructing the
stars photometric behaviour on short timescales. The overall trend
indicates a peak around 1905 and, in total, a decline until 1915.

Additional plate data from HDAP (Figure
\ref{licu_VarC}; filled red diamonds) further confirm the indication from
the Mount Wilson data (\cite{1953ApJ...118..353H}; black dots) of a broad
maximum between 1900 and 1915. The lack of data around 1905 and before 1899
prevents us from putting stronger constrains on the maximum.

More regular time coverage of Var~C is available starting in 1918 when the
star showed a brightness of approximately 17.5~mag in B consistently in
the Mount Wilson and HADP data. Until the mid 1930s, the star remained at
almost constant luminosity apart from some smaller variations of less than
half a magnitude. Subsequently, the luminosity of Var~C started to increase
over more than ten years until it reached its first established prominent
maximum. 
With a gap in the data coverage present between 1941 and 1945, the 
slope of the light curve towards the maximum is not sampled very well. 
This maximum around 1947 with a maximum brightness in B of about 15.4~mag 
lasted at least three years. The luminosity started to decline around 1950. 
The sparse data coverage also makes it hard to determine 
the minimum value with good accuracy, which is probably almost as faint 
as before rising.
 
Minor increases of up to one magnitude are seen around 1957 and
1963. Especially the later rise and decrease is very narrow with almost one
magnitude in less than one month. A broader and less luminous minor maximum
around 1967 can be assumed.

With the exception of smaller variations between 1970 and 1982, Var~C stayed
constant at a B brightness of $\approx$17~mag. Since most of the values at
this phase are photographic or photoelectric measurements, there might be a
zeropoint or colourterm offset. When taking this into account, it is probably
more significant for judging the variability to only look at variations within
each set of data values separately. Since all data of one data set were very
likely processed in the same way, variations in the data values can be trusted
to be real.

The next prominent maximum occurred around 1986 where the luminosity again
rose to about 15.4~mag for about four years. The rapid rise -- in particular
in the R band (compare Figure 1 from \citet{1988AA...203..306H}) -- was
followed by a  steep decline in luminosity, which occurred at the end of
1986. It can be seen well
in the TLS plate data (fig. \ref{licu_VarC} red crosses) but is also present
in the data by \citet{1988AA...203..306H} (fig. \ref{licu_VarC} blue
diamonds) and \citet{1995AJ....110.2715M} (fig. \ref{licu_VarC} yellow
four-ray star). Within 120 days, variations of about one magnitude are seen.

Around 1990 the luminosity in B of Var~C was almost back to a value before the
rise. During the next 20~years minor and/or steeper increases and declines
took place, namely in 1992, 1996, end of 2002, and 2007. The maximum around
2002/2003 was nearly as bright as the two previous prominent maxima, but a
much steeper increase and a decrease is present in this one.

In September 2009 the luminosity in B was 17.56~mag. It 
brightened in October 2011 and reached a new maximum value of 
15.87~mag on 1 September 2013 \citep{2013ATel..5362....1H} and has stayed 
at this new maximum light level ever since. That Var~C
stayed at maximum light during this time and did not pass through
two close and short maxima, is even more evident in the V light 
curve, that has additional measurements (see Figure \ref{licu_BV} 
middle panel).

%__________________________________________________________________

\section{Spectra of Var~C}

The first spectra of Var~C have already been reported in
\citet{1953ApJ...118..353H}. Spectra were taken between November 1946 and
November 1951. The first two spectra were taken in November 1946 and in August
1947, when Var~C was in a phase of maximum light (see Figure
\ref{licu_VarC}). The spectra show \CaII\ K ($\lambda$3933) and
\CaII\ H ($\lambda$3968) in absorption. Both lines have nearly equal
strengths. According to \citet{1953ApJ...118..353H}, this is ``[...] indicating
that the spectral class is at least later than F0.'' In the first of these two
spectra, \Hg\ and \Hd\ are weakly present in absorption, while the second one
shows no hydrogen lines.  The next spectra were taken during descending
light. In 1949 the spectra showed faint \Hb\ emission but no \CaII\ K or
\CaII\ H lines were visible. The spectra taken from August to October 1951
show a much stronger \Hb\ line and an \Hg\ line, both in emission. \CaII\ K
and \CaII\ H were seen in absorption. In November 1951, \Hb\ and \Hg\ were
still seen in emission, but no \CaII\ K and \CaII\ H lines were visible.

Based on their spectrum obtained in September 1973,
\citet{1975SvAL....1...30S} reported that Var~C was displaying a bright
\Ha\ emission. \citet{1975ApJ...200..426H} recorded two spectra taken in
October and November 1974 that showed several hydrogen, \HeI\, \FeII, and
[\FeII] lines in emission, as well as \CaII\ K in absorption. She stated that
Var~C is an $\eta$~Carina-like object.

\citet{1978ApJ...219..445H} compared a spectrum taken two years later in
August 1976 with the two spectra taken previously. The hydrogen and
\HeI\ emission lines had become weaker and some of the \FeII\ and [\FeII] lines
were no longer seen.

The spectra taken in November 1983 by \citet{1985ApJ...290..542K} showed \Ha,
\Hg, and numerous \FeII\ lines in emission. All these lines showed pronounced
P-Cygni profiles. Strong \MgII\ $\lambda$4481 was seen in
absorption. \citet{1985ApJ...290..542K} suggested a temperature for Var~C at
that time of only slightly less than $<$10000~K.

Several optical spectra of Var~C taken between November 1985 and September
1987, as well as one UV spectrum (August 1986), were presented in
\citet{1988AA...203..306H}. The spectra of Var~C taken in 1985 --
corresponding to a maximum phase of Var~C -- showed only \Ha\ and \Hb\ in
emission. Several strong lines of ionised metals including \FeII\ were seen
in absorption. \citet{1988AA...203..306H} stated that the spectrum resembles
that of an early F-type supergiant (F0Ia--F5Ia). They estimated a surface
temperature of $\approx$7500~K. 
The optical spectra taken in 1986 showed weaker metallic lines, 
indicating---together with its ratio of the \CaII\ K and \CaII\ H 
lines---a higher surface temperature of up to 9000~K and a spectral 
type of about A2 to A3 \citep{1988AA...203..306H}.
    
The UV spectrum showed no
emission lines, but \FeII\ lines in absorption underlined the
estimation of Var~C being an A-type star at that time. In the high-resolution
spectra taken in September 1987 \FeII\ and other metallic lines were seen in
absorption, as well as several \FeII-lines in emission. \Ha, \Hb, and
\Hg\ showed P-Cygni profiles.

\citet{1996AA...314..131S} presented optical spectra taken in November 1991,
October 1992, and December 1992. An additional UV spectrum was taken in July
1992. The 1992 spectra were described to show weak
\HeI\ lines in absorption corresponding to a temperature $>$10000~K at that
time. A comparison made between the spectra from 1987
\citep[presented in][]{1988AA...203..306H}, 1991, and 1992 shows that the
hydrogen and \FeII\ emission lines were getting stronger. Both optical and UV
spectra were described to be consistent with a post-maximum phase of Var~C.

A spectrum taken in December 1993 (\citet{1995AJ....110.2715M},
\citet{1996ApJ...469..629M}, and \citet{2007AJ....134.2474M}) showed
pronounced hydrogen lines in emission and several \FeII\ and [\FeII] in
emission. \HeI\ is weakly seen.
Spectra taken in December 2004 and January 2005 by \citet{2006AA...458..225V}
showed strong hydrogen lines in emission. Several \FeII\ and [\FeII] lines
were seen in emission. A spectral type of B[e] was given.

\citet{2012AA...541A.146C} reported on two spectra taken on 30 November 2003
and 29 September through 2 October 2010. The first one showed \Hb\ in emission
and numerous metallic absorption lines. The spectrum was compared to a
spectrum of the F0-F5Ia$^{+}$ hypergiant B324. The spectrum taken in 2010
showed \Ha, \Hb, and \Hg\ in emission. Several \HeI\ lines were also seen in
emission, as was \FeII. The emission lines showed P-Cygni profiles. Var~C became hotter from
2003 to 2010. As seen from the light curve, the luminosity
of Var~C decreased during this time.

Two SAO RAS spectra taken on 13 November 2004 and 22 October 2008 are
presented in Figure \ref{spec_varc}. The spectrum recorded in November 2004
shows \Ha, \Hb, and \Hg\ in emission. Several weak \FeII\ lines are
present in emission. In the spectrum from October 2008, \Ha\ and \Hb\ are seen
in emission. Several \HeI\ and some \FeII\ lines are very present in
emission. Comparing this development with the slope of the light curve
shows that Var~C faded between the end of 2004 and the end of 2008 of about
0.7~mag. This is consistent with the spectra and indicates that the star
is entering a hot phase.

A spectrum of Var~C taken 3 October 2010 \citep{2014ApJ...782L..21H}
shows Balmer lines in emission. Several \HeI\ and numerous \FeII\ and
[\FeII] lines are also seen in emission. Pronounced P-Cygni profiles are
present in the spectrum. With respect to the light curve, the spectrum was
taken after a minimum, with Var~C increasing its luminosity, but still
representing a hot star.

A low-resolution spectrum taken on 18 September 2013 by
\citet{2013ATel.5403....1R} is described to show hydrogen lines in emission.

Two spectra taken on 5 October 2013 and 1-2 November 2013 were reported by
\citet{2013ATel.5538....1V}. The spectra were described to show \FeII\ lines
in emission, as well as [\FeII] and strong hydrogen lines. \HeI\ was seen very
weakly in absorption. P-Cygni profiles were present in the hydrogen lines and
the majority of the \FeII\ lines.

Another spectrum taken on 7 October 2013 by \citet{2014ApJ...782L..21H}
showed much weaker hydrogen lines in emission,
\CaII\ and \MgII\ in absorption, and \FeII\ in emission with P-Cygni
profiles. The spectrum is classified as late A-type. This spectral
type was still confirmed with an LBT IR spectrum taken January 2014.

Table \ref{Spec_param} summarises the optical spectral observations of
Var~C\footnote{Based on a misidentification during service,} observing
the spectrum reported by \citet{2011BSRSL..80..356B} taken in September 2007
was not that of Var~C, but of a different star. Where not 
already given in the literature, a spectral type has been estimated based upon
the description of the spectrum in the literature. The criteria for
classifications were derived from spectral descriptions given by
\cite{1987clst.book.....J}.

%__________________________________________________________________

\begin{figure}
   \centering
   \includegraphics[width=\columnwidth]{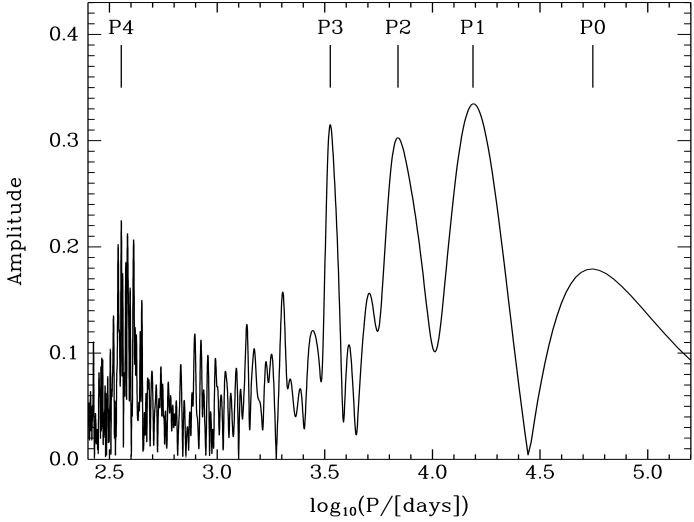}
   \caption{Power spectrum derived from the B magnitude values presented
     in the upper panel of Figure \ref{licu_spec}. The highest peak (P1)
     corresponds to a period of 42.3~years. The values for the other peaks
are given in Table \ref{fourier_spec_tab}.}
   \label{fourier_spec_2}
\end{figure}

\begin{figure}
   \centering
   \includegraphics[width=\columnwidth]{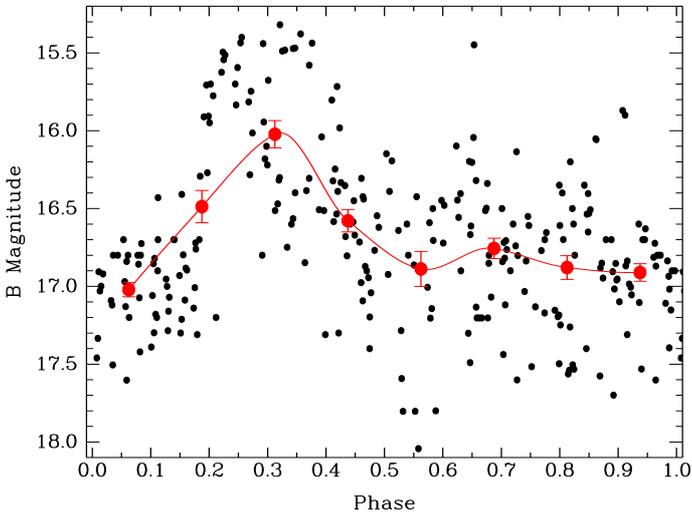}
   \caption{Phase diagram using a frequency of $\nu=6.477\times
     10^{-5}$\,1/d (P1$\approx$42.3~years). The data has been binned (red
     dots) and the binned data fitted with an akima spline (red
     line). A zero point offset in time of 3000 days was applied.}
   \label{phase_p1}
\end{figure}

\section{Discussion}

\subsection{Periodicity}

While several semi-periodic structures are known to exist in the light
curves of LBVs on timescales of about ten years,
the periods are less stable than in Cepheids
\citep{2001AA...366..508V}, for example. The extremely long light curve of
Var~C that we established manifests an excellent unique data set to make a
reliable analysis for even long-term trends and periods.
A first inspection of the light curve of Var~C reveals two pronounced maxima.
The first maximum occurred around 1947 and a second one around 1986.
To investigate the long scale variability in B of Var~C, we analysed the
periodicity using a Fourier analysis performed with Period04
\citep{2005CoAst.146...53L}. Data values from \citet{1975ApJS...29..303V} were
excluded because of the relatively large magnitude uncertainties of 0.7~mag, as
was one data point from WIYN for the same reason. The B data values
were averaged by month in order to avoid being sensitive to small scale
variations (below a month). A light curve of these averaged data is presented
in Figure \ref{licu_spec} (upper panel).

\begin{table}
\caption{Peak values corresponding to Figure \ref{fourier_spec_2}.}
\label{fourier_spec_tab}
\centering
\begin{tabular}{l c c c c}
\hline\hline
Peak & $\log _{10}$(P/[d]) & Period P   & Frequency $\nu$       & Amplitude \\
     &                    & [years]    & [1/d]                 &           \\
\hline
P0   & 4.745        & 152.2 & $1.799 \times 10^{-5}$ & 0.179 \\
P1   & 4.189        & 42.3  & $6.477 \times 10^{-5}$ & 0.335 \\
P2   & 3.841        & 19.0  & $1.444 \times 10^{-4}$ & 0.303 \\
P3   & 3.526        & 9.2   & $2.980 \times 10^{-4}$ & 0.315 \\
P4   & 2.555        & 1     & $2.786 \times 10^{-3}$ & 0.225 \\
\hline
\end{tabular}
\end{table}

Figure \ref{fourier_spec_2} shows the power spectrum derived from the B data
values. A bright peak is seen at P1=15440 days=42.3~years. Two more
peaks are present at P2=6926 days=19.0~years and P3=3356 days=9.2~years.
The peaks around P4 corresponding to a period of approximately one year are
most probably due to the sampling of the data. The very broad and low
amplitude peak (P0) with the highest period is due to the apparent overall
increase in the luminosity. Such a secular trend of brightening is also seen
in the light curve of other LBVs like $\eta$~Car. Table
\ref{fourier_spec_tab} lists the parameters of the main peaks labelled in Figure
\ref{fourier_spec_2}.

As seen from Table \ref{fourier_spec_tab} periods, P1 and P2 are roughly
multiples of period P3. The amplitudes of P1, P2, and P3 are also in the same
order of magnitude. This could mean that the 40-year period is not real,
which would then favour the twenty- or ten-year period.

To gain a better understanding of the structures found in the power spectrum,
synthetic light curves were produced. Therefore, synthetic data modelling the
main features of Var~C's light curve were generated. Subsequently, Period04
was used on these data.

One of these synthetic light curves was generated with one data point per
month at only six months per year, while no data
points were set the other half of the year. This was done to reproduce that
M33 is only observable during roughly half a year each year from the northern
hemisphere. Using this setup, we were able to reproduce the broad bump
consisting of several peaks around P4. When adding data points for the other
six months, this peaks vanish. This indicates that the peak around P4 is
indeed produced by the sampling of the data.

As another test, data points were added to the light curve of Var~C by
averaging neighbouring data points and adding the averaged value as a new
data point in between them. This method was applied twice, so that a light
curve with two times and another with four times the original number of data
points were created. Both times Period04 found the same peaks (P1, P2, P3)
with P1 having the highest amplitude, even though this still does not
necessarily mean that the 40-year period is real.

The phase diagram corresponding to the maximum peak at a frequency of
$\nu=6.477\times 10^{-5}$~per day (corresponding to period P1=15440
days=42.3~years) is given in Figure \ref{phase_p1}. The data has been
binned into eight bins and these binned data has been fitted with
an akima spline. A zero point offset in time of 3000 days was applied.

At least one prominent maximum is seen in the phase diagram. A second, smaller
maximum at approximately half of the period might be assumed, but data
scattering is much larger and renders it quite uncertain. This minor maximum
is represented in Figure \ref{fourier_spec_2} by peak P1. It still might be that
the period is only approximately 18-20~years with the maximum amplitude
not always equally strong. Another interpretation would be that the period is
$\approx$40~years with a major maximum and a minor maximum. Assuming the 
last major maximum was around 1986, the next strong maximum phase of Var~C 
should occur around 2028$\pm$3. Nevertheless,(semi-)periodic behaviour on 
large time scales is seen in the light curve.

Some uncertainties arise from the fact that the data are from various
telescopes and that transformations had to be made between different magnitude
systems ($m_{pg}$, B). Assuming all data of one dataset have been processed in
the same way, variations within such a dataset are more significant and give
information about variability on smaller scales.

\citet{2001AA...366..508V} found two frequencies (both roughly a year;
P$_0$=371.4 days; P$_1$=305 days or 475 days) present in AG~Car and
superimposed on each other, resulting in a beat cycle of about 20~years. This
looks similar to the 20-year period found in Var~C's light curve (if assuming
maxima of different amplitudes).

To check whether the periodicity in Var~C's light curve on this long time
scale might only be the result of a superposition of periodicities on smaller
timescales, we also looked at the small variations of Var~C. For this we used
the data from the TLS plate scans to have a consistent data set.
In addition we also used V band data from \citet{2001AJ....121..870M} (images
taken between September 1996 and October 1997) and \citet{2006MNRAS.370.1429S}
(images taken between September 2000 and November 2003), covering a range of
six years to search for periodicity.

Even though variabilities on smaller scales are obviously present in the light
curve of Var~C, no clear period around a year was found. Such small
and intermediate scale periodicities are not even seen
when looking only at separate datasets and at datasets with the highest data
coverage. This means either that the data coverage is still not good enough to
find these small scale periodicities or that the long-term periodicity found in
Var~C is not just a beat cycle resulting from the superposition of two or
more frequencies, as in AG Car. This could mean that the periodicity of
Var~C is caused by a different mechanism than the long-term periodicity of AG
Car.

Also no periodicity  of days was found in Var~C's light curve as in UIT301. \citet{2000MNRAS.311..698S} reported the M33 LBV UIT301
(also named B416) to show a periodicity of 8.26 days. Even though
\citet{2004BaltA..13..156S} stated that this is the half of the orbital period
and therefore light variations are due to a close interacting binary rather
than to intrinsic variability in luminosity.

With an amplitude of approximately 1.5-2~mag for the major peak (and 0.5 mag
for a minor peak -- if present) and with an assumed length of 42~years, Var~C
can be classified as a strong-active S~Dor member with a long S~Dor (L-SD)
cycle, as defined by \citet{2001AA...366..508V}.
Comparing Var~C's light curve with the light curves of other 
strong-active S~Dor members given in \citet{2001AA...366..508V}, it is 
seen that, induced by sparse data coverage, the light curves are 
quite patchy.
Time approximations for the L-SD durations indicate variabilities of
decades, but no periodicity on these timescales can obviously be seen from
the light curves. Most light curves show secular trends of de- or increasing
light (e.g. R110). Some light curves show a single maximum (e.g. R116).

So far, regular--periodic--variations have not been known to be a common
feature of LBVs. With the really long-term light curves, which have come up
more recently, the first cases are being observed right now.
\cite{2014ATel.6295....1W} have reported on the LMC LBV
R71 to show periodic variability on a timescale of about 40 years. More
recently, the maxima have appeared more frequently and with a larger
amplitude. This is similar to what was observed for the LBV Var~B in M33
between 1930 and 1950, when \cite{1953ApJ...118..353H} found three maxima with
steadily increasing amplitudes.

Periodicity on a timescale of several decades is unusually long. Common
mechanisms causing periodic light variations, such as the $\kappa$-mechanism,
occur on much smaller timescales (days up to a few months). So far, no intrinsic
stellar mechanism has been known to cause such long-term periodicity.

Recently, \citet{2010AJ....139.2600K} have found a 40-year cyclic variation in
HD~5980, which is a multiple system consisting of a close binary (two
Wolf-Rayet type stars orbiting each other with a period of 19.3 days) and a
third O-type star component. The authors suggest that the observed slow
variations due to changes in the radius were superimposed by strong and
short-duration eruptions caused by the binary companion.

The similarity in the length of the periods found in Var~C and HD~5980 might
lead to the question of whether periodicity caused by interaction in a binary
system is a general phenomenon and thus whether Var~C might also consist of
more than one component. So far, no hints of binarity have been reported for
Var~C. Also the available high-resolution spectra do not show evidence of
any binarity.

Nevertheless, the photometric variations seen in HD~5980 and Var~C appear to
be quite different. The light curve of HD~5980 (see
\citet{2010AJ....139.2600K} Figure 1) shows a sudden, short eruption after a
slow brightening over several years. This peak is very steep and narrow. In
contrast to that, Var~C's first maximum shows a relatively slow rise and
fall. The second maximum appears slightly steeper and narrower than the first
one, but is still much broader than the peak seen in HD~5980. Both of Var~C's
maxima last at least a few years.

If Var~C is indeed a single star, the similarity of the period might just be
coincidental. In that case a completely different underlying mechanism has to
be responsible for the (semi-)pe\-riodicity of Var~C.

\subsection{Temperature variations}

\addtocounter{table}{1}

\begin{table*}
\caption{Temperature estimations for Var~C. Depending on the value of (B-V)$_0$ Equation \ref{4b}, \ref{4c}, \ref{4d}, or \ref{4e} was used.}
\label{temperatures}
\centering
\begin{tabular}{l c c c c c c c c c c}
\hline\hline
& & & \multicolumn{2}{c}{E(B-V)=0.10} & \multicolumn{2}{c}{E(B-V)=0.17} & \multicolumn{2}{c}{E(B-V)=0.24} & & Observatory/\\
Date            & B     & V     & (B-V)$_0$ & T/K       & (B-V)$_0$ & T/K       & (B-V)$_0$ & T/K & T/K           & Reference\\
\hline
Aug. 1963       & 16.74 & 16.36 & 0.28  & 7306  & 0.21  & 7723  & 0.14  & 8291    & 7800$\pm$500          & TLS plate\\
Sept. 1963      & 16.01 & 15.75 & 0.16  & 8098  & 0.09  & 8792  & 0.02  & 9546    & 8800$\pm$700          & TLS plate\\
Sept. 1976      & 17.12 & 17.17 & -0.15 & 14541 & -0.22 & 20743 & -0.29 & 29591   & 21600$\pm$8000        & 1\\
Oct. 1977       & 16.98 & 16.98 & -0.10 & 11282 & -0.17 & 16095 & -0.24 & 22959   & 16800$\pm$6200        & 2\\
Nov. 1980       & 17.20 & 17.21 & -0.11 & 11870 & -0.18 & 16932 & -0.25 & 24155   & 17700$\pm$6500        & 3\\
Dec. 1981       & 17.33 & 17.23 & 0.00  & 9772  & -0.07 & 9689  & -0.14 & 13822   & 11100$\pm$2700        & 3\\
Aug. 1982       & 17.24 & 17.15 & -0.01 & 7146  & -0.08 & 10193 & -0.15 & 14541   & 10600$\pm$3900        & 3\\
Oct. 1982       & 16.57 & 16.48 & -0.01 & 7146  & -0.08 & 10193 & -0.15 & 14541   & 10600$\pm$3900        & 3\\
Sept./Oct. 1986 & 16.43 & 16.35 & -0.02 & 7518  & -0.09 & 10724 & -0.16 & 15298   & 11200$\pm$4100        & 4\\
Nov. 1986       & 16.12 & 15.75 & 0.27  & 7364  & 0.20  & 7784  & 0.13  & 8389    & 7800$\pm$500          & 5\\
Sept. 1987      & 16.12 & 15.68 & 0.34  & 6967  & 0.27  & 7364  & 0.20  & 7784    & 7400$\pm$400          & 5\\
Aug./Oct. 1988  & 16.51 & 16.39 & 0.02  & 9546  & -0.05 & 8754  & -0.12 & 12488   & 10300$\pm$2200        & 6\\
Sept. 1992      & 16.30 & 16.11 & 0.09  & 8792  & 0.02  & 9546  & -0.05 & 8754    & 9000$\pm$500          & 7\\
Dec. 1992       & 16.54 & 16.34 & 0.10  & 8690  & 0.03  & 9434  & -0.04 & 8321    & 8800$\pm$600          & 7\\
Jan. 1993       & 16.39 & 16.21 & 0.08  & 8896  & 0.01  & 9658  & -0.06 & 9210    & 9300$\pm$400          & 7\\
Sept. 1996      & 17.15 & 16.18 & 0.87  & 5139  & 0.80  & 5326  & 0.73  & 5520    & 5300$\pm$200          & DIRECT\\
Oct. 1996       & 17.25 & 16.81 & 0.34  & 6967  & 0.27  & 7364  & 0.20  & 7784    & 7400$\pm$400          & DIRECT\\
Aug. 1997       & 17.82 & 16.90 & 0.82  & 5272  & 0.75  & 5464  & 0.68  & 5663    & 5500$\pm$200          & DIRECT\\
Sept. 1997      & 17.59 & 16.87 & 0.62  & 5839  & 0.55  & 6052  & 0.48  & 6236    & 6000$\pm$200          & DIRECT\\
Oct. 1997       & 17.80 & 16.79 & 0.91  & 5035  & 0.84  & 5218  & 0.77  & 5408    & 5200$\pm$200          & DIRECT\\
Aug. 1998       & 17.80 & 17.23 & 0.47  & 6285  & 0.40  & 6644  & 0.33  & 7022    & 6700$\pm$400          & DIRECT\\
Nov. 1998       & 18.04 & 17.90 & 0.04  & 9324  & -0.03 & 7909  & -0.10 & 11282   & 9500$\pm$1800         & DIRECT\\
Oct. 1999       & 17.20 & 17.08 & 0.02  & 9546  & -0.05 & 8754  & -0.12 & 12488   & 10300$\pm$2200        & 8\\
Nov. 1999       & 17.18 & 17.07 & 0.01  & 9658  & -0.06 & 9210  & -0.13 & 13138   & 10700$\pm$2500        & 8\\
Oct. 2000       & 16.65 & 16.53 & 0.02  & 9546  & -0.05 & 8754  & -0.12 & 12488   & 10300$\pm$2200        & NOAO\\
Sept. 2001      & 16.45 & 16.33 & 0.02  & 9546  & -0.05 & 8754  & -0.12 & 12488   & 10300$\pm$2200        & NOAO\\
Aug. 2002       & 16.67 & 15.83 & 0.74  & 5492  & 0.67  & 5692  & 0.60  & 5899    & 5700$\pm$200          & WIYN\\
Sept. 2002      & 16.94 & 15.88 & 0.96  & 4908  & 0.89  & 5087  & 0.82  & 5272    & 5100$\pm$200          & WIYN\\
Oct. 2002       & 16.43 & 15.86 & 0.47  & 6285  & 0.40  & 6644  & 0.33  & 7022    & 6700$\pm$400          & WIYN\\
Nov. 2002       & 16.27 & 15.84 & 0.33  & 7022  & 0.26  & 7423  & 0.19  & 7818    & 7400$\pm$400          & WIYN\\
Jan. 2003       & 17.05 & 15.81 & 1.14  & 4476  & 1.07  & 4639  & 1.00  & 4808    & 4600$\pm$200          & WIYN\\
Sept. 2003      & 16.51 & 15.87 & 0.54  & 6083  & 0.47  & 6285  & 0.40  & 6644    & 6300$\pm$300          & WIYN\\
Oct. 2003       & 16.50 & 15.56 & 0.84  & 5218  & 0.77  & 5408  & 0.70  & 5605    & 5400$\pm$200          & WIYN\\
Nov. 2003       & 16.34 & 15.87 & 0.37  & 6803  & 0.30  & 7191  & 0.23  & 7601    & 7200$\pm$400          & WIYN\\
Dec. 2003       & 16.80 & 15.74 & 0.96  & 4908  & 0.89  & 5087  & 0.82  & 5272    & 5100$\pm$200          & WIYN\\
Dec. 2004       & 16.50 & 16.40 & 0.00  & 9772  & -0.07 & 9689  & -0.14 & 13822   & 11100$\pm$2700        & 9\\
Jul. 2005       & 16.90 & 16.26 & 0.54  & 6083  & 0.47  & 6285  & 0.40  & 6644    & 6300$\pm$300          & WIYN\\
Sept. 2005      & 16.60 & 16.60 & -0.10 & 11282 & -0.17 & 16095 & -0.24 & 22959   & 16800$\pm$6200        & 9\\
Oct. 2007       & 16.81 & 16.86 & -0.15 & 14541 & -0.22 & 20743 & -0.29 & 29591   & 21600$\pm$8000        & CoSAI\\
Oct. 2007       & 17.02 & 16.94 & -0.02 & 7518  & -0.09 & 10724 & -0.16 & 15298   & 11200$\pm$4100        & TLS CCD\\
Jan. 2008       & 17.17 & 17.08 & -0.01 & 7146  & -0.08 & 10193 & -0.15 & 14541   & 10600$\pm$3900        & TLS CCD\\
Oct. 2008       & 17.15 & 17.07 & -0.02 & 7518  & -0.09 & 10724 & -0.16 & 15298   & 11200$\pm$4100        & TLS CCD\\
Jan. 2009       & 17.19 & 17.10 & -0.01 & 7146  & -0.08 & 10193 & -0.15 & 14541   & 10600$\pm$3900        & TLS CCD\\
Feb. 2009       & 17.25 & 17.20 & -0.05 & 8754  & -0.12 & 12488 & -0.19 & 17814   & 13000$\pm$4800        & TLS CCD\\
Oct. 2009       & 17.53 & 17.47 & -0.04 & 8321  & -0.11 & 11870 & -0.18 & 16932   & 12400$\pm$4600        & TLS CCD\\
Nov. 2009       & 17.23 & 16.86 & 0.27  & 7364  & 0.20  & 7784  & 0.13  & 8389    & 7900$\pm$500          & CoSAI\\
Feb. 2010       & 17.53 & 17.51 & -0.08 & 10193 & -0.15 & 14541 & -0.22 & 20743   & 15200$\pm$5600        & TLS CCD\\
Feb. 2011       & 16.40 & 16.21 & 0.09  & 8792  & 0.02  & 9546  & -0.05 & 8754    & 9000$\pm$500          & TLS CCD\\
Aug. 2011       & 16.13 & 15.81 & 0.22  & 7662  & 0.15  & 8194  & 0.08  & 8896    & 8300$\pm$600          & TLS CCD\\
Sept. 2011      & 15.98 & 15.79 & 0.09  & 8792  & 0.02  & 9546  & -0.05 & 8754    & 9000$\pm$500          & TLS CCD\\
Oct. 2011       & 16.06 & 15.81 & 0.15  & 8194  & 0.08  & 8896  & 0.01  & 9658    & 8900$\pm$700          & TLS CCD\\
Sep. 2013       & 15.87 & 15.66 & 0.11  & 8588  & 0.04  & 9324  & -0.03 & 7909    & 8600$\pm$700          & 10\\
Nov. 2013       & 15.9  & 15.7  & 0.1   & 8690  & 0.03  & 9434  & -0.04 & 8321  & 8800$\pm$600            & 11\\
\hline
\end{tabular}
\tablebib{(1) \citet{1978ApJ...219..445H};
(2) \citet{1978ApJ...221L..73H};
(3) \citet{1984ApJ...278..124H};
(4) \citet{1995AJ....110.2715M};
(5) \citet{1988AA...203..306H};
(6) \citet{1991AJ....101.1663W};
(7) \citet{1996AA...314..131S};
(8) \citet{2001AJ....122.2477M};
(9) \citet{2006AA...458..225V};
(10) \citet{2013ATel..5362....1H};
(11) \citet{2013ATel.5538....1V}}
\end{table*}

By converting the photometrical colours of Var~C into temperatures, a more
physical interpretation can be made. For this we used Equations (4b) through
(4e) from \citet{1993AJ....106.1471P} (see also references within).

For $(B-V)_0 < 0.0$ we used Equation (4b):
\begin{equation}
   \label{4b}
   \log(T_{eff})=3.832-2.204(B-V)_0,
\end{equation}
\noindent
for $0.0 \leq (B-V)_0 < 0.2$ Equation (4c) was applied:
\begin{equation}
   \label{4c}
   \log(T_{eff})=3.990-0.510(B-V)_0.
\end{equation}
\noindent
For $0.2 \leq (B-V)_0 < 0.5$ Equation (4d) was used,
\begin{equation}
   \label{4d}
   \log(T_{eff})=3.960-0.344(B-V)_0,
\end{equation}
\noindent
and for $0.5 \leq (B-V)_0$ Equation (4e) was applied:
\begin{equation}
   \label{4e}
   \log(T_{eff})=3.904-0.222(B-V)_0.
\end{equation}

\begin{figure*}
   \centering
   \includegraphics[width=\textwidth]{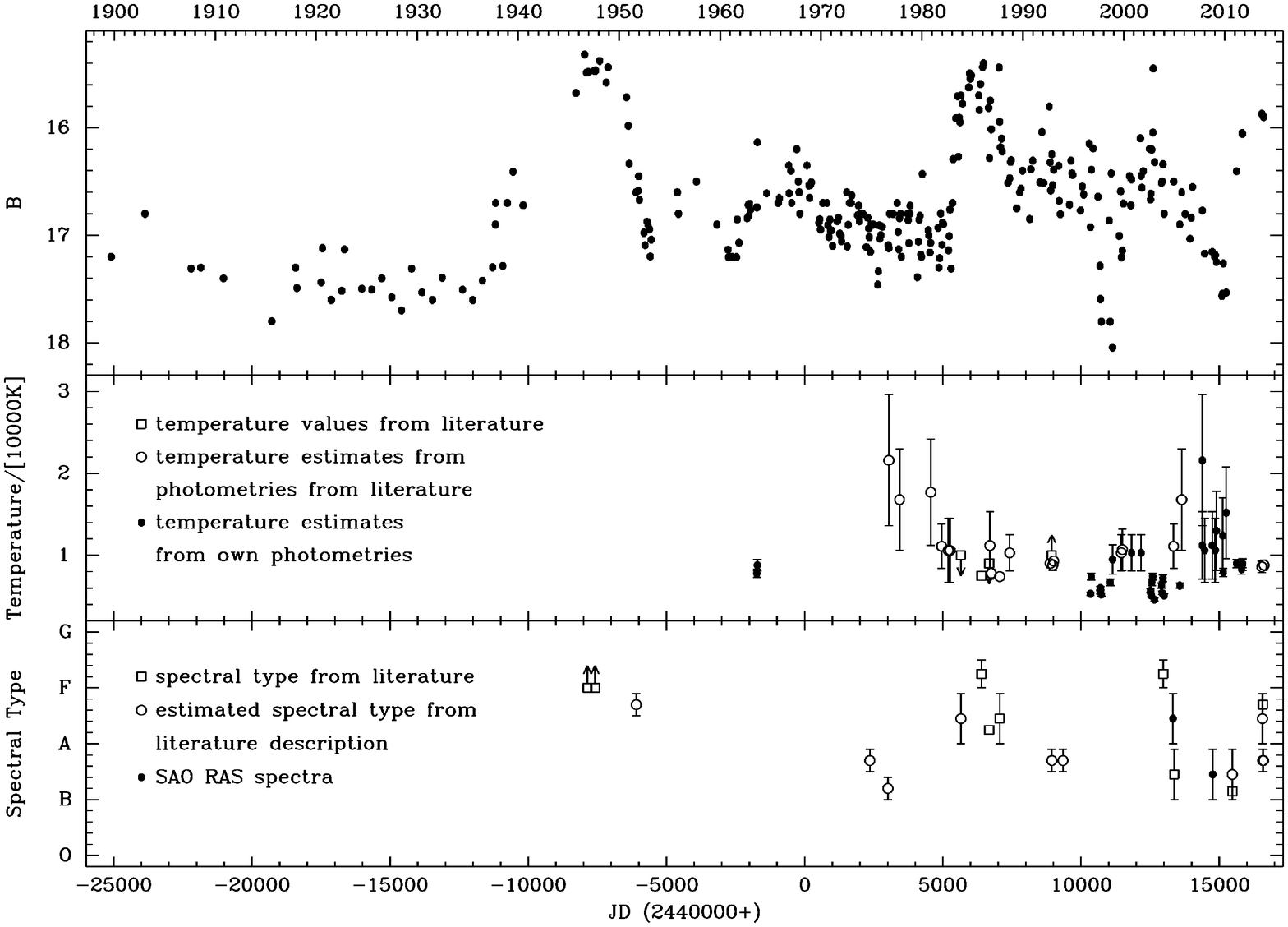}
   \caption{B light curve of Var~C (upper panel). In comparison to
     Figure \ref{licu_VarC}, the B values here have been averaged by month. The
     corresponding temperatures and spectral types are given in the middle and
     lower panels, respectively.}
   \label{licu_spec}
\end{figure*}

\citet{1993AJ....106.1471P} used these equations to calculate $T_{eff}$ for
stars with only photometric values available, hence for stars without
any further spectral classification. According to them, Equation \ref{4b} is
not valid for supergiants. Supergiants would be cooler (a few 1000~K,
depending on the spectral type) than dwarfs of the same spectral
type. Therefore, the estimated temperature would be to high, if the star was a
supergiant.

Furthermore, the calculation of the temperature strongly depends on the
assumed reddening of the star. Therefore, we determined the reddening from an
U-B versus B-V colour-colour diagram using stars in the surrounding of
Var~C. Since Var~C only has a few neighbouring stars, the determination of
reddening is rendered somewhat uncertain. A mean value of 0.17$\pm$0.07 was
found. This is in good agreement with a reddening value for Var~C from
e.g. \citet{2006AA...458..225V}, who used A$_V$=0.6, which equals
E(B-V)=0.19. We calculated temperatures for three different values of E(B-V)
of 0.10, 0.17, and 0.24.

Because of these uncertainties (validity of equations, reddening),
these temperature calculations can only give a rough estimate. Nevertheless,
they can be used to trace the general development of Var~C's temperature
curve. Table \ref{temperatures} lists the calculated temperatures. The
calculations were done for colours derived from own photometries and
from photometries from the literature. The resulting temperature curve is
given in Figure \ref{licu_spec} (middle panel).

Figure \ref{licu_spec} illustrates the connection between light variations and
changes in temperature, hence in spectral type. The upper panel shows the B
light curve of Var~C. In contrast to the light curve presented in Figure
\ref{licu_VarC}, the B values here were averaged by month. These are also the
B values used for analysing the periodicity. The middle panel
presents the corresponding temperatures. 
Literature values and
temperature estimated from photometric values are plotted. Finally, the lower
panel shows the spectral types of Var~C at different times. Values are either taken
directly from the literature or are estimated spectral types from the
description of a spectrum in the literature (see also Table \ref{Spec_param}).

Even though the coverage with spectral data is quite patchy, it is seen that
during maximum light, the star resembles an A- or F-type star. During phases of
minimum light an O- or B-type star is seen. Since changes in the spectral type
are caused by changes in the temperature, this trend is also seen in the
temperature curve. For example, during the maximum around 1986, temperatures around
10000~K are seen, while temperatures above 16000~K
are present just before this maximum. Also the decrease in luminosity between 2003 and 2010 is
accompanied by a rise in the temperature curve, while lower temperatures are measured again in the beginning maximum
after 2010.

\begin{figure*}
   \centering
   \includegraphics[width=\textwidth]{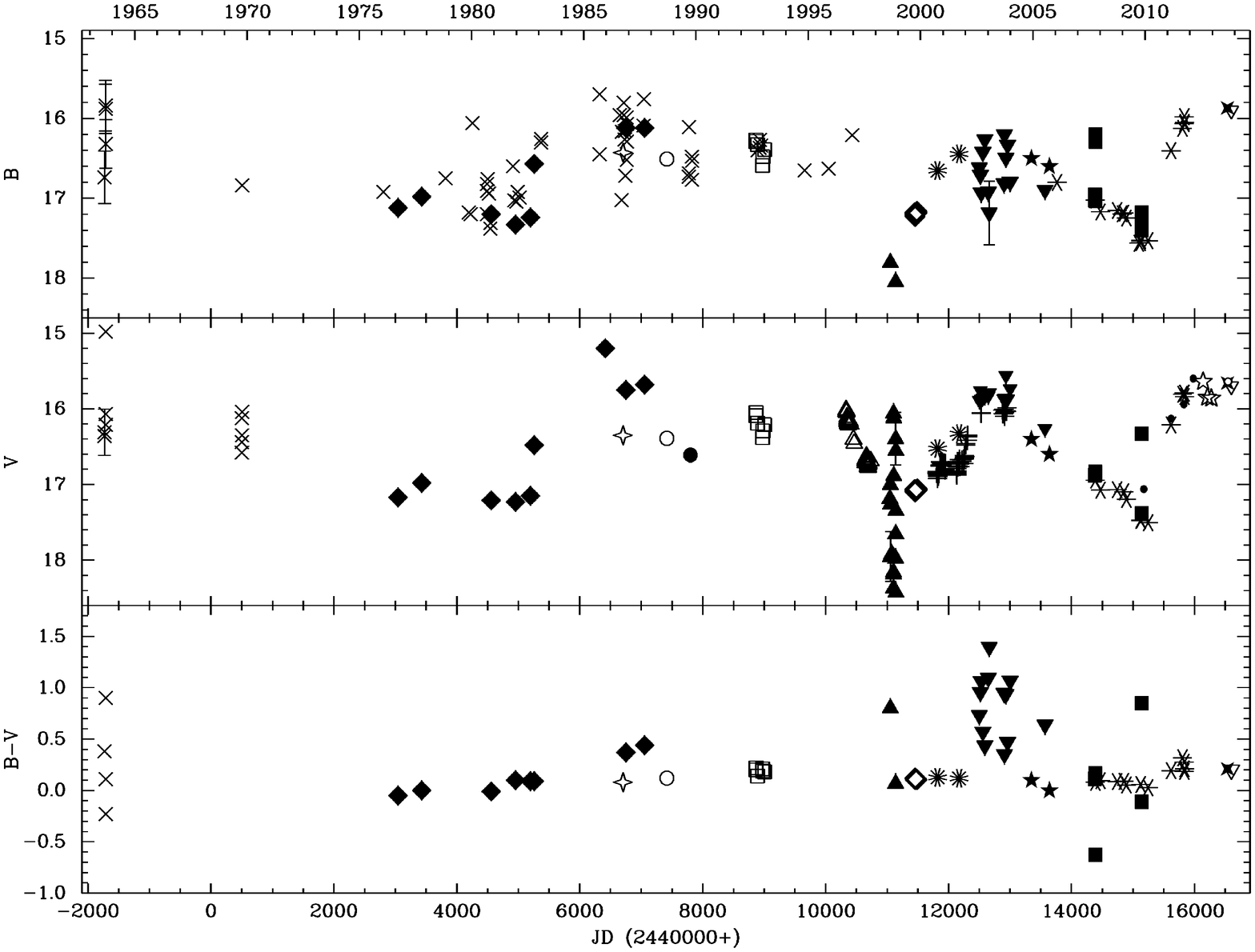}
   \caption{B and V light curves and B$-$V colour curve of Var~C. Crosses: TLS
     plate data; filled upwards triangles: DIRECT; stars: NOAO; filled
     downwards triangles: WIYN; stars with six rays: TLS CCD data; filled
     squares: COoSAI; filled diamonds: \citet{1978ApJ...219..445H},
     \citet{1978ApJ...221L..73H}, \citet{1984ApJ...278..124H}, and
     \citet{1988AA...203..306H}; open four-ray stars:
     \citet{1995AJ....110.2715M}; open circles: \citet{1991AJ....101.1663W};
     filled circle: \citet{Spiller92}; open squares:
     \citet{1996AA...314..131S}; open upwards triangles: \citet{2001AJ....121..870M};
     open diamonds: \citet{2001AJ....122.2477M}; plus signs:
     \citet{2006MNRAS.370.1429S}; filled five-ray stars:
     \citet{2006AA...458..225V}; filled small dots:
     \citet{2013ATel.5403....1R}; open small dot: Roberto Viotti, Franco
     Montagni (private communication); open five five-ray stars: John Martin
     (private communication); filled four-ray stars:
     \citet{2013ATel..5362....1H}; and open downwards triangles:
     \citet{2013ATel.5538....1V}.
     Only data sets containing V band data were used in these plots. Only
     errors larger than 0.3~mag are shown.}
   \label{licu_BV}
\end{figure*}

Some additional V band data were available for Var~C. Therefore, a V
light curve is also presented in the middle panel of Figure \ref{licu_BV}. For a
comparison, the B light curve is shown in the upper panel of the same
figure. The corresponding B$-$V colour curve is given in the lower panel.

As for changes in the spectral type, changes in the colour (here B$-$V)
also represent changes in the temperatures of a star. Lower B$-$V values indicate
redder, hence cooler stars. As seen from Figure \ref{licu_BV}, the
B$-$V colour curve also shows that Var~C is redder and cooler when it is in maximum
light.

All these described trends in the temperature and light variations -- cooler
when in minimum light and hotter while in maximum light -- indicate S-Dor
behaviour, the variability intrinsic to LBVs. A detailed SED fitting,
however, is not possible because for most epochs, no multi-wavelength observations
have been recorded.

%______________________________________________________________

\section{Conclusions}

We investigated the long-term photometric and spectral behaviour of the LBV
Var~C in M33. We classified Var~C as a strong-active S~Dor member with a long
S~Dor (L-SD) cycle.
A long-term (semi-)periodicity with a period of 42.3~years is detected
with the chance of the real period being 19.0~years or even 9.2~years. If a
long-term periodicity of approximately 42~years is present in the B light
curve of Var~C, the next maximum light should occur around
2028$\pm$3. However, developments since 2010 have shown that Var~C is already
entering a phase of maximum light 15~years earlier, which is not consistent
with the longest possible period, but is within the uncertainty that matching both the shorter periods.

Var~C has been in maximum light for more than one year now
\citep{2013ATel..5362....1H}. Even though Var~C is approximately 0.5 mag
fainter than during the two prominent maxima in 1947 and 1986, the slope of
the rise and the duration so far fits the previous two bright maxima,
indicating a comparable eruptive state. Var~C's spectral, colour, and temperature curves, together with its
corresponding light curves, show variations that indicate an S~Dor variability.

Several measurements were reported during the tracing of the current changes
in the light curve of Var~C. This shows that even though Var~C is a long known
LBV, it is still necessary and important to trace its behaviour further in
order to fully understand the various features of LBVs.

As recently shown for the LBV R71 in the LMC \citep{2014ATel.6295....1W}, only
by compiling a light curve over $\sim$100~years is it possible to detect
photometric variation on long timescales. These features' (as in the case
of R71) two broad maxima at 1914 and 1939 give rise to physical events in the
star, which are currently unknown and far from being understood. The brightness
of R71 has increased during the last five~years and reached the highest
maximum ever reported \citep{2014ATel.6295....1W}. In addition,
the R71 photometric behaviour seems like what we found for
Var~C. Indications of regular change between bright and dim peaks with similar
maxima and minima (in magnitudes) are visible in R71, as for Var~C. R71 shows
broad and overlayed narrow maxima that appear to have some (but not
necessarily strict) periodicity, similar to the structures in the light
curve of Var~C that we report in this paper. Also, an overlayed secular trend
towards higher luminosity seems to be present, again very similar to the one
in Var~C.

Another feature of the R71 light curve is the trend toward the 
duration of the maxima being shorter with increasing amplitudes 
\citep{2014ATel.6295....1W}. This may also be present in the 
light curve of Var~C. It also may explain the
rather small amplitude of the 1905 maximum of Var~C indicated in our light
curve. The increasing time interval between outbursts of R71
\citep{2014ATel.6295....1W} does not appear to be present in the light curve of
Var~C. Therefore, Var~C may not (yet) be in an evolutionary phase of accelerated
changes \citep[e.g.][]{2014MNRAS.439.2917M}. This is also supported by the
absence of \CaII\ emission lines (and therefore a dense envelope), as reported
for R71 (\citet{2009AJ....138.1003B},
\citet{2012CBET.3192....1G},\citet{2013msao.confE.168M}). The historic light
curve of R71 appears to be rather more strongly modulated than that of Var~C,
and especially the recent outburst of R71 (\citet{2012CBET.3192....1G},
\citet{2013msao.confE.168M}) is much more extreme than the one of Var~C
\citep{2013ApJ...773...46H}.

Based on the similarity to R71, one may speculate that the semi-periodicity plus
the secular brightening trend in the light curve of Var~C is an indication
that the evolutionary state of both stars is going to change, and the current,
rather unexpected brightening of Var~C is an indication of a coming large
eruption as in R71. If this should happen to Var~C soon, it would give rise to
a new phenomenon in LBVs: large eruptions following L-S~Dor variability
are overlayed on a secular brightening trend on timescales of 100~years and may be
linked to accelerated evolution towards a supernova. A monitoring of the
subsequent development of the current bright state of Var~C is therefore much
needed. Clearly, a hunt for more historic plates will be very fruitful.

%______________________________________________________________

\begin{acknowledgements}
     We thank Marko R\"oder and Bringfried Stecklum for service observations
      with the 2m-Alfred-Jensch-telescope of the Th\"uringer Landessternwarte
      (TLS) Tautenburg. Thanks go to Otmar Stahl for helpful comments on the
      classification of our spectrum of Var~C and Gloria Koenigsberger for
      discussions on the variability of massive stars. We thank Roberta
      M. Humphreys and Kris Davidson for inspiring discussions and helpful
      comments. Special thanks go to Roberta M. Humphreys for her
      help with line identification. We are grateful to Artie P. Hatzes, John
      Martin,
      and Ren\'e Hudic for comments and Chris Evans for a spectrum of
      Var~C. B. Burggraf is thankful for support by a stipend from the
      ``Wilhelm and G\"unter Esser foundation''. O. Sholukhova and A. Zharova
      are grateful for RFBR grant No.~13-02-00885, the grant ``Leading
      Scientific Schools of Russia'', and grant No.~14-50-00043 of the
      Russian Scientific Foundation.
      This work made use of the HDAP which was
      produced at the Landessternwarte Heidelberg-K\"onigstuhl under grant
      No. 00.071.2005 of the Klaus-Tschira-Foundation. We thank
      the referee for detailed comments.  

\end{acknowledgements}

\bibliographystyle{aa}
\bibliography{Var_C_V2.bib}

\begin{thebibliography}{70}
\expandafter\ifx\csname natexlab\endcsname\relax\def\natexlab#1{#1}\fi

\bibitem[{{Afanasiev} \& {Moiseev}(2005)}]{2005AstL...31..194A}
{Afanasiev}, V.~L. \& {Moiseev}, A.~V. 2005, Astronomy Letters, 31, 194

\bibitem[{{Bertin} \& {Arnouts}(1996)}]{1996A&AS..117..393B}
{Bertin}, E. \& {Arnouts}, S. 1996, \aaps, 117, 393

\bibitem[{{Bonanos} {et~al.}(2009){Bonanos}, {Massa}, {Sewilo}, {Lennon},
  {Panagia}, {Smith}, {Meixner}, {Babler}, {Bracker}, {Meade}, {Gordon},
  {Hora}, {Indebetouw}, \& {Whitney}}]{2009AJ....138.1003B}
{Bonanos}, A.~Z., {Massa}, D.~L., {Sewilo}, M., {et~al.} 2009, \aj, 138, 1003

\bibitem[{{Brunzendorf} \& {Meusinger}(1999)}]{1999AAS..139..141B}
{Brunzendorf}, J. \& {Meusinger}, H. 1999, \aaps, 139, 141

\bibitem[{{Burggraf} {et~al.}(2011){Burggraf}, {Weis}, {Bomans}, \&
  {Henze}}]{2011BSRSL..80..356B}
{Burggraf}, B., {Weis}, K., {Bomans}, D.~J., \& {Henze}, M. 2011, Bulletin de
  la Societe Royale des Sciences de Liege, 80, 356

\bibitem[{{Clark} {et~al.}(2012){Clark}, {Castro}, {Garcia}, {Herrero},
  {Najarro}, {Negueruela}, {Ritchie}, \& {Smith}}]{2012AA...541A.146C}
{Clark}, J.~S., {Castro}, N., {Garcia}, M., {et~al.} 2012, \aap, 541, A146

\bibitem[{{Conti}(1984)}]{1984IAUS..105..233C}
{Conti}, P.~S. 1984, in IAU Symposium, Vol. 105, Observational Tests of the
  Stellar Evolution Theory, ed. A.~{Maeder} \& A.~{Renzini}, 233

\bibitem[{{Demleitner} {et~al.}(2001){Demleitner}, {Accomazzi}, {Eichhorn},
  {Grant}, {Kurtz}, \& {Murray}}]{2001ASPC..238..321D}
{Demleitner}, M., {Accomazzi}, A., {Eichhorn}, G., {et~al.} 2001, in
  Astronomical Society of the Pacific Conference Series, Vol. 238, Astronomical
  Data Analysis Software and Systems X, ed. F.~R. {Harnden}, Jr., F.~A.
  {Primini}, \& H.~E. {Payne}, 321

\bibitem[{{Dolphin}(2000)}]{2000PASP..112.1383D}
{Dolphin}, A.~E. 2000, \pasp, 112, 1383

\bibitem[{{Fabrika} \& {Sholukhova}(1995)}]{1995ApSS.226..229F}
{Fabrika}, S. \& {Sholukhova}, O. 1995, \apss, 226, 229

\bibitem[{{Gamen} {et~al.}(2012){Gamen}, {Walborn}, {Morrell}, {Barba}, \&
  {Fernandez Lajus}}]{2012CBET.3192....1G}
{Gamen}, R., {Walborn}, N., {Morrell}, N., {Barba}, R., \& {Fernandez Lajus},
  E. 2012, Central Bureau Electronic Telegrams, 3192, 1

\bibitem[{{Henze} {et~al.}(2008){Henze}, {Meusinger}, \&
  {Pietsch}}]{2008AA...477...67H}
{Henze}, M., {Meusinger}, H., \& {Pietsch}, W. 2008, \aap, 477, 67

\bibitem[{{Hubble} \& {Sandage}(1953)}]{1953ApJ...118..353H}
{Hubble}, E. \& {Sandage}, A. 1953, \apj, 118, 353

\bibitem[{{Humphreys}(1975)}]{1975ApJ...200..426H}
{Humphreys}, R.~M. 1975, \apj, 200, 426

\bibitem[{{Humphreys}(1978)}]{1978ApJ...219..445H}
{Humphreys}, R.~M. 1978, \apj, 219, 445

\bibitem[{{Humphreys} {et~al.}(1984){Humphreys}, {Blaha}, {D'Odorico}, {Gull},
  \& {Benvenuti}}]{1984ApJ...278..124H}
{Humphreys}, R.~M., {Blaha}, C., {D'Odorico}, S., {Gull}, T.~R., \&
  {Benvenuti}, P. 1984, \apj, 278, 124

\bibitem[{{Humphreys} \& {Davidson}(1994)}]{1994PASP..106.1025H}
{Humphreys}, R.~M. \& {Davidson}, K. 1994, \pasp, 106, 1025

\bibitem[{{Humphreys} {et~al.}(2014){Humphreys}, {Davidson}, {Gordon}, {Weis},
  {Burggraf}, {Bomans}, \& {Martin}}]{2014ApJ...782L..21H}
{Humphreys}, R.~M., {Davidson}, K., {Gordon}, M.~S., {et~al.} 2014, \apjl, 782,
  L21

\bibitem[{{Humphreys} {et~al.}(2013{\natexlab{a}}){Humphreys}, {Davidson},
  {Grammer}, {Kneeland}, {Martin}, {Weis}, \& {Burggraf}}]{2013ApJ...773...46H}
{Humphreys}, R.~M., {Davidson}, K., {Grammer}, S., {et~al.} 2013{\natexlab{a}},
  \apj, 773, 46

\bibitem[{{Humphreys} {et~al.}(1988){Humphreys}, {Leitherer}, {Stahl}, {Wolf},
  \& {Zickgraf}}]{1988AA...203..306H}
{Humphreys}, R.~M., {Leitherer}, C., {Stahl}, O., {Wolf}, B., \& {Zickgraf},
  F.-J. 1988, \aap, 203, 306

\bibitem[{{Humphreys} \& {Warner}(1978)}]{1978ApJ...221L..73H}
{Humphreys}, R.~M. \& {Warner}, J.~W. 1978, \apjl, 221, L73

\bibitem[{{Humphreys} {et~al.}(2013{\natexlab{b}}){Humphreys}, {Weis},
  {Burggraf}, {Bomans}, {Martin}, \& {Meusinger}}]{2013ATel..5362....1H}
{Humphreys}, R.~M., {Weis}, K., {Burggraf}, B., {et~al.} 2013{\natexlab{b}},
  The Astronomer's Telegram, 5362, 1

\bibitem[{{Ivanov} {et~al.}(1993){Ivanov}, {Freedman}, \&
  {Madore}}]{1993ApJS...89...85I}
{Ivanov}, G.~R., {Freedman}, W.~L., \& {Madore}, B.~F. 1993, \apjs, 89, 85

\bibitem[{{Jaschek} \& {Jaschek}(1987)}]{1987clst.book.....J}
{Jaschek}, C. \& {Jaschek}, M. 1987, {The classification of stars}, ed.
  M.~Jaschek, C. \&~Jaschek

\bibitem[{{Kaluzny} {et~al.}(1998){Kaluzny}, {Stanek}, {Krockenberger},
  {Sasselov}, {Tonry}, \& {Mateo}}]{1998AJ....115.1016K}
{Kaluzny}, J., {Stanek}, K.~Z., {Krockenberger}, M., {et~al.} 1998, \aj, 115,
  1016

\bibitem[{{Kenyon} \& {Gallagher}(1985)}]{1985ApJ...290..542K}
{Kenyon}, S.~J. \& {Gallagher}, III, J.~S. 1985, \apj, 290, 542

\bibitem[{{Koenigsberger} {et~al.}(2010){Koenigsberger}, {Georgiev}, {Hillier},
  {Morrell}, {Barb{\'a}}, \& {Gamen}}]{2010AJ....139.2600K}
{Koenigsberger}, G., {Georgiev}, L., {Hillier}, D.~J., {et~al.} 2010, \aj, 139,
  2600

\bibitem[{{Kurtev} {et~al.}(1999){Kurtev}, {Corral}, \&
  {Georgiev}}]{1999AA...349..796K}
{Kurtev}, R.~G., {Corral}, L.~J., \& {Georgiev}, L. 1999, \aap, 349, 796

\bibitem[{{Lenz} \& {Breger}(2005)}]{2005CoAst.146...53L}
{Lenz}, P. \& {Breger}, M. 2005, Communications in Asteroseismology, 146, 53

\bibitem[{{Lovas} \& {Zsoldos}(1988)}]{1988IBVS.3193....1L}
{Lovas}, M. \& {Zsoldos}, E. 1988, Information Bulletin on Variable Stars,
  3193, 1

\bibitem[{{Macri} {et~al.}(2001){Macri}, {Stanek}, {Sasselov}, {Krockenberger},
  \& {Kaluzny}}]{2001AJ....121..870M}
{Macri}, L.~M., {Stanek}, K.~Z., {Sasselov}, D.~D., {Krockenberger}, M., \&
  {Kaluzny}, J. 2001, \aj, 121, 870

\bibitem[{{Maeder}(1983)}]{1983AA...120..113M}
{Maeder}, A. 1983, \aap, 120, 113

\bibitem[{{Massey} {et~al.}(1995){Massey}, {Armandroff}, {Pyke}, {Patel}, \&
  {Wilson}}]{1995AJ....110.2715M}
{Massey}, P., {Armandroff}, T.~E., {Pyke}, R., {Patel}, K., \& {Wilson}, C.~D.
  1995, \aj, 110, 2715

\bibitem[{{Massey} {et~al.}(1996){Massey}, {Bianchi}, {Hutchings}, \&
  {Stecher}}]{1996ApJ...469..629M}
{Massey}, P., {Bianchi}, L., {Hutchings}, J.~B., \& {Stecher}, T.~P. 1996,
  \apj, 469, 629

\bibitem[{{Massey} {et~al.}(2001){Massey}, {Hodge}, {Holmes}, {Jacoby}, {King},
  {Olsen}, {Saha}, \& {Smith}}]{2001AAS...19913005M}
{Massey}, P., {Hodge}, P.~W., {Holmes}, S., {et~al.} 2001, Bulletin of the
  American Astronomical Society, 33, 1496

\bibitem[{{Massey} {et~al.}(2002){Massey}, {Hodge}, {Holmes}, {Jacoby}, {King},
  {Olsen}, {Smith}, \& {Saha}}]{2002AAS...20110407M}
{Massey}, P., {Hodge}, P.~W., {Holmes}, S., {et~al.} 2002, Bulletin of the
  American Astronomical Society, 34, 1272

\bibitem[{{Massey} {et~al.}(2007){Massey}, {McNeill}, {Olsen}, {Hodge},
  {Blaha}, {Jacoby}, {Smith}, \& {Strong}}]{2007AJ....134.2474M}
{Massey}, P., {McNeill}, R.~T., {Olsen}, K.~A.~G., {et~al.} 2007, \aj, 134,
  2474

\bibitem[{{Massey} {et~al.}(2006){Massey}, {Olsen}, {Hodge}, {Strong},
  {Jacoby}, {Schlingman}, \& {Smith}}]{2006AJ....131.2478M}
{Massey}, P., {Olsen}, K.~A.~G., {Hodge}, P.~W., {et~al.} 2006, \aj, 131, 2478

\bibitem[{{Mehner} {et~al.}(2013){Mehner}, {Baade}, {Rivinius}, {Lennon},
  {Martayan}, {Stahl}, \& {Stefl}}]{2013msao.confE.168M}
{Mehner}, A., {Baade}, D., {Rivinius}, T., {et~al.} 2013, in Massive Stars:
  From alpha to Omega, 168

\bibitem[{{Meusinger} {et~al.}(2010){Meusinger}, {Henze}, {Birkle}, {Pietsch},
  {Williams}, {Hatzidimitriou}, {Nesci}, {Mandel}, {Ertel}, {Hinze}, \&
  {Berthold}}]{2010AA...512A...1M}
{Meusinger}, H., {Henze}, M., {Birkle}, K., {et~al.} 2010, \aap, 512, A1

\bibitem[{{Meynet} \& {Maeder}(2005)}]{2005AA...429..581M}
{Meynet}, G. \& {Maeder}, A. 2005, \aap, 429, 581

\bibitem[{Mighell(1998)}]{ccdcap}
Mighell, K.~J. 1998, CCDCAP: CCD Circular Aperture Photometry

\bibitem[{{Mochejska} {et~al.}(2001){Mochejska}, {Kaluzny}, {Stanek},
  {Sasselov}, \& {Szentgyorgyi}}]{2001AJ....122.2477M}
{Mochejska}, B.~J., {Kaluzny}, J., {Stanek}, K.~Z., {Sasselov}, D.~D., \&
  {Szentgyorgyi}, A.~H. 2001, \aj, 122, 2477

\bibitem[{{Moriya} {et~al.}(2014){Moriya}, {Maeda}, {Taddia}, {Sollerman},
  {Blinnikov}, \& {Sorokina}}]{2014MNRAS.439.2917M}
{Moriya}, T.~J., {Maeda}, K., {Taddia}, F., {et~al.} 2014, \mnras, 439, 2917

\bibitem[{{Nijland}(1901)}]{1901AN....154..413N}
{Nijland}, A.~A. 1901, Astronomische Nachrichten, 154, 413

\bibitem[{{Parker} \& {Garmany}(1993)}]{1993AJ....106.1471P}
{Parker}, J.~W. \& {Garmany}, C.~D. 1993, \aj, 106, 1471

\bibitem[{{Pellerin} \& {Macri}(2011)}]{2011ApJS..193...26P}
{Pellerin}, A. \& {Macri}, L.~M. 2011, \apjs, 193, 26

\bibitem[{{Rosino} \& {Bianchini}(1973)}]{1973AA....22..453R}
{Rosino}, L. \& {Bianchini}, A. 1973, \aap, 22, 453

\bibitem[{{Scowcroft} {et~al.}(2009){Scowcroft}, {Bersier}, {Mould}, \&
  {Wood}}]{2009MNRAS.396.1287S}
{Scowcroft}, V., {Bersier}, D., {Mould}, J.~R., \& {Wood}, P.~R. 2009, \mnras,
  396, 1287

\bibitem[{{Sharov}(1973)}]{1973PZ.....19....3S}
{Sharov}, A.~S. 1973, Peremennye Zvezdy, 19, 3

\bibitem[{{Sharov}(1990)}]{1990SvA....34..364S}
{Sharov}, A.~S. 1990, Soviet Astronomy, 34, 364

\bibitem[{{Sharov} {et~al.}(1975){Sharov}, {Esipov}, \&
  {Lyutyi}}]{1975SvAL....1...30S}
{Sharov}, A.~S., {Esipov}, V.~F., \& {Lyutyi}, V.~M. 1975, Soviet Astronomy
  Letters, 1, 30

\bibitem[{{Shemmer} {et~al.}(2000){Shemmer}, {Leibowitz}, \&
  {Szkody}}]{2000MNRAS.311..698S}
{Shemmer}, O., {Leibowitz}, E.~M., \& {Szkody}, P. 2000, \mnras, 311, 698

\bibitem[{{Sholukhova} {et~al.}(2007){Sholukhova}, {Abolmasov}, {Fabrika}, \&
  {Afanasiev}}]{2007ASPC..361..491S}
{Sholukhova}, O., {Abolmasov}, P., {Fabrika}, S., \& {Afanasiev}, V. 2007, 361,
  491

\bibitem[{{Sholukhova} {et~al.}(2004){Sholukhova}, {Fabrika}, {Roth}, \&
  {Becker}}]{2004BaltA..13..156S}
{Sholukhova}, O., {Fabrika}, S., {Roth}, M., \& {Becker}, T. 2004, Baltic
  Astronomy, 13, 156

\bibitem[{{Shporer} \& {Mazeh}(2006)}]{2006MNRAS.370.1429S}
{Shporer}, A. \& {Mazeh}, T. 2006, \mnras, 370, 1429

\bibitem[{{Spiller}(1992)}]{Spiller92}
{Spiller}, F.~O. 1992, PhD thesis, Universit\"at Heidelberg

\bibitem[{{Szeifert} {et~al.}(1996){Szeifert}, {Humphreys}, {Davidson},
  {Jones}, {Stahl}, {Wolf}, \& {Zickgraf}}]{1996AA...314..131S}
{Szeifert}, T., {Humphreys}, R.~M., {Davidson}, K., {et~al.} 1996, \aap, 314,
  131

\bibitem[{{Valeev} {et~al.}(2013){Valeev}, {Fabrika}, \&
  {Sholukhova}}]{2013ATel.5538....1V}
{Valeev}, A., {Fabrika}, S., \& {Sholukhova}, O. 2013, The Astronomer's
  Telegram, 5538, 1

\bibitem[{{van den Bergh} {et~al.}(1975){van den Bergh}, {Herbst}, \&
  {Kowal}}]{1975ApJS...29..303V}
{van den Bergh}, S., {Herbst}, E., \& {Kowal}, C.~T. 1975, \apjs, 29, 303

\bibitem[{{van Genderen}(2001)}]{2001AA...366..508V}
{van Genderen}, A.~M. 2001, \aap, 366, 508

\bibitem[{{van Genderen} {et~al.}(1997){van Genderen}, {Sterken}, \& {de
  Groot}}]{1997A&A...318...81V}
{van Genderen}, A.~M., {Sterken}, C., \& {de Groot}, M. 1997, \aap, 318, 81

\bibitem[{{Viotti} {et~al.}(2013){Viotti}, {Rossi}, {Montagni}, \&
  {Gualandi}}]{2013ATel.5403....1R}
{Viotti}, R., {Rossi}, C., {Montagni}, F., \& {Gualandi}, R. 2013, The
  Astronomer's Telegram, 5403, 1

\bibitem[{{Viotti} {et~al.}(2006){Viotti}, {Rossi}, {Polcaro}, {Montagni},
  {Gualandi}, \& {Norci}}]{2006AA...458..225V}
{Viotti}, R.~F., {Rossi}, C., {Polcaro}, V.~F., {et~al.} 2006, \aap, 458, 225

\bibitem[{{Walborn} {et~al.}(2014){Walborn}, {Gamen}, {Barba}, \&
  {Morrell}}]{2014ATel.6295....1W}
{Walborn}, N.~R., {Gamen}, R.~C., {Barba}, R.~H., \& {Morrell}, N.~I. 2014, The
  Astronomer's Telegram, 6295, 1

\bibitem[{{Weaver} {et~al.}(1977){Weaver}, {McCray}, {Castor}, {Shapiro}, \&
  {Moore}}]{1977ApJ...218..377W}
{Weaver}, R., {McCray}, R., {Castor}, J., {Shapiro}, P., \& {Moore}, R. 1977,
  \apj, 218, 377

\bibitem[{{Weis}(2011)}]{2011BSRSL..80..440W}
{Weis}, K. 2011, Bulletin de la Societe Royale des Sciences de Liege, 80, 440

\bibitem[{{Weis} \& {Bomans}(2005)}]{2005AA...429L..13W}
{Weis}, K. \& {Bomans}, D.~J. 2005, \aap, 429, L13

\bibitem[{{Wilson}(1991)}]{1991AJ....101.1663W}
{Wilson}, C.~D. 1991, \aj, 101, 1663

\bibitem[{{Zharova} \& {Sholukhova}(2004)}]{2004CoAst.145...28Z}
{Zharova}, A. \& {Sholukhova}, O. 2004, Communications in Asteroseismology,
  145, 28

\end{thebibliography}

\end{document}